\documentclass[smallextended, sn-mathphys-num]{sn-jnl}
 
\usepackage{graphicx}%
\usepackage{multirow}%
\usepackage{amsmath,amssymb,amsfonts}%
\usepackage{amsthm}%
\usepackage{mathrsfs}%
\usepackage[title]{appendix}%
\usepackage{xcolor}%
\usepackage{textcomp}%
\usepackage{manyfoot}%
\usepackage{booktabs}%
\usepackage{algorithm}%
\usepackage{algorithmicx}%
\usepackage{algpseudocode}%
\usepackage{listings}%
\usepackage{adjustbox}
\usepackage{caption}
\usepackage{subcaption}

\usepackage{amssymb}
\usepackage{lipsum}
\usepackage{todonotes}
\usepackage{enumitem}

\usepackage{float}
\usepackage[utf8]{inputenc}
\usepackage{csquotes}
\graphicspath{{figures/}}
\PassOptionsToPackage{hyphens}{url}\usepackage{hyperref}
\usepackage{array}

\usepackage{multirow}
\usepackage{xcolor}
\usepackage{comment}
\usepackage{balance}
\usepackage{hyphenat}
\usepackage{wrapfig}

\expandafter\def\expandafter\UrlBreaks\expandafter{\UrlBreaks
  \do\a\do\b\do\c\do\d\do\e\do\f\do\g\do\h\do\i\do\j%
  \do\k\do\l\do\m\do\n\do\o\do\p\do\q\do\r\do\s\do\t%
  \do\u\do\v\do\w\do\x\do\y\do\z\do\A\do\B\do\C\do\D%
  \do\E\do\F\do\G\do\H\do\I\do\J\do\K\do\L\do\M\do\N%
  \do\O\do\P\do\Q\do\R\do\S\do\T\do\U\do\V\do\W\do\X%
  \do\Y\do\Z\do\*\do\-\do\~\do\'\do\"\do\-}%

\newcommand{\hl}[1]{\textcolor{red}{#1}}

\theoremstyle{thmstyleone}%
\theoremstyle{thmstyletwo}%

\theoremstyle{thmstylethree}%

\begin{document}

\title[AI Techniques in the Microservices Life-Cycle]{AI Techniques in the Microservices Life-Cycle: \\ A Systematic Mapping Study}


\author*[1,2]{\fnm{Sergio} \sur{Moreschini}}\email{sergio.moreschini@tuni.fi}

\author*[3]{\fnm{Shahrzad} \sur{Pour}}\email{shmp@dtu.dk}

\author*[4]{\fnm{Ivan} \sur{Lanese}}\email{ivan.lanese@gmail.com}

\author[5]{\fnm{Daniel} \sur{Balouek}}
\equalcont{These authors contributed equally to this work.}

\author[7]{\fnm{Justus} \sur{Bogner}}
\equalcont{These authors contributed equally to this work.}

\author[2]{\fnm{Xiaozhou} \sur{Li}}
\equalcont{These authors contributed equally to this work.}

\author[8]{\fnm{Fabiano} \sur{Pecorelli}}
\equalcont{These authors contributed equally to this work.}

\author[9]{\fnm{Jacopo} \sur{Soldani}}
\equalcont{These authors contributed equally to this work.}

\author[10]{\fnm{Eddy} \sur{Truyen}}
\equalcont{These authors contributed equally to this work.}

\author[2,1]{\fnm{Davide} \sur{Taibi}}

\affil*[1]{\orgname{Tampere University}, \orgaddress{\city{Tampere}, \country{Finland}}}

\affil[2]{\orgname{University of Oulu}, \orgaddress{\city{Oulu}, \country{Finland}}}

\affil[3]{\orgdiv{DTU Compute}, \orgname{Technical University of Denmark},
\orgaddress{\city{Lyngby}, \country{Denmark}}}

\affil[4]{\orgdiv{OLAS team}, \orgname{University of Bologna/INRIA}, \orgaddress{\city{Bologna}, \country{Italy}}}

\affil[5]{\orgname{Inria, LS2N Laboratory, IMT Atlantique}, \orgaddress{\city{Nantes}, \country{France}}}

\affil[7]{\orgname{Vrije Universiteit Amsterdam}, \orgaddress{\city{Amsterdam}, \country{The Netherlands}}}

\affil[8]{ \orgname{Pegaso Digital University}, \orgaddress{\city{Naples}, \country{Italy}}}

\affil[9]{\orgdiv{Department of Computer Science}, \orgname{University of Pisa}, \orgaddress{
\city{Pisa}, 
\country{Italy}}}

\affil[10]{\orgdiv{DistriNet}, \orgname{KU Leuven}, \orgaddress{
\city{Leuven}, 
\country{Belgium}}}


\abstract{
The use of AI in microservices (MSs) is an emerging field as indicated by a substantial number of surveys. However these surveys focus on a specific problem using specific AI techniques, therefore not fully capturing the growth of research and the rise and disappearance of trends. In our systematic mapping study, we take an exhaustive approach to reveal all possible connections between the use of AI techniques for improving any quality attribute (QA) of MSs during the DevOps phases. Our results include 16 research themes that connect to the intersection of particular QAs, AI domains and DevOps phases. Moreover by mapping identified future research challenges and relevant industry domains, we can show that many studies aim to deliver prototypes to be automated at a later stage, aiming at providing exploitable products in a number of key industry domains.
 }




\maketitle

\section{Introduction}
\label{Introduction}
Microservices (MSs)~\cite{Dragoni2017} is a popular architectural style for distributed applications, which originated from service-oriented computing and pushed the concept of modularity much further than its ancestors. As such, an MS-based system consists of small, loosely coupled, and possibly heterogeneous services, which can be deployed, updated, and scaled independently. This is often supported by executing individual services in containers, such as the ones provided by Docker~\cite{docker}. Containerization ensures that services can be easily moved or duplicated. As a result, MSs can provide high flexibility, scalability, and evolvability. However, these advantages come at a price: an MS system often comprises many fine-grained services, which may interact according to complex patterns. Mastering this complexity is challenging.

In recent years, Artificial Intelligence (AI) in general and Machine Learning (ML) in particular have attracted considerable interest from research and practice~\cite{Liu2018}. As a result, AI techniques have been applied in various application areas, and software engineering is no exception~\cite{Kotti2023}. 
In particular, AI has been applied in numerous works to support the development and operations of MSs. However, to the best of our knowledge, the role of AI for MSs is still unclear, and no holistic secondary studies analyzed the adoption of AI techniques for MSs in an exhaustive manner.  

This paper focuses on the use of AI techniques to solve challenges or improve the quality of MS-based systems (AI4MS), e.g., regarding design, development, and operation. We would like to understand how and why AI techniques are used within  MS Architecture (MSA) and its life-cycle, which AI approaches are used, in which industry domains, and which challenges are still open for future research. 
To this end, we report on what is being said in the literature on the topic, by providing a sort of \enquote{snapshot} of the state-of-the-art on AI4MS. We indeed
performed a systematic mapping study (SMS)~\cite{PETERSEN20151}, investigating how the publication landscape evolved over the years and including 269 peer-reviewed papers published between 2017 and up to and including 2023.
The aim of our SMS is to overview the \textit{when}, \textit{where}, \textit{why}, and \textit{how} of AI4MS, while also shedding light on open research challenges in the field. Among other things, we study how the number of AI4MS publications has evolved over the years, in which industry domains AI4MS is used, which quality attributes (QAs) it improves in which DevOps phases, and by means of which AI techniques.

To extract data according to taxonomies as uniform and unbiased as possible, we reused established classifications whenever possible. For instance, for AI techniques we used the classification in~\cite{AIWatch}, and for improved QAs we referred to the ISO 25010:2011 (SQuaRE) standard~\cite{InternationalOrganizationForStandardization2011}. For the phases of the software life-cycle in which an approach is used, we referred to the DevOps life-cycle~\cite{Yildirim2019}.

The results of our work can inform researchers about the relationship between AI and MSs, with a focus on how modern AI techniques are used to improve MS systems. Such information can be used by researchers to take informed decisions on AI-based techniques to consider when designing future MSAs and to investigate valuable open challenges. Also, a refined understanding of which QAs are improved by using AI and in which DevOps phases can be useful to practitioners interested in enhancing their MS systems.
The main contribution of this paper is a report on the state of the art concerning AI techniques to support MSs.  


{\bf Paper structure:}
Section~\ref{Background} introduces the background information on MS and the related works about AI4MS. Our research method is described in Section~\ref{Methodology}. Section~\ref{Results} provides the results for the five research questions individually, while Section~\ref{sec:sa_qa_ai} refines the results of RQs 2-4 via a multidimensional analysis. Section \ref{Discussion} presents the main discussion points and implications originating from the analysis. The possible threats to the study's validity are in Section~\ref{Threats}. 
Section~\ref{Conclusion} concludes the paper.  
\section{Background on Microservices and Related Work}
\label{Background}

\subsection{Microservices}
As a reimagination of the service-oriented architectures (SOAs)~\cite{Erl2005} approach, 
MSA started to rise in popularity around 2014~\cite{Dragoni2017,Fowler2014,Newman2015}.
However, the MSA architectural style was already used by several companies, such as Netflix~\cite{Wang2013}.
Today, MSA is fairly popular in industry, e.g., 37\% of developers surveyed by JetBrains in 2022 responded they were using MSs\footnote{\url{https://www.jetbrains.com/lp/devecosystem-2022/microservices/}}.
MSs are also a popular research topic today, with a substantial number of publications each year.
Google Scholar\footnote{\url{https://scholar.google.com}, queried on 2024-09-30} 
reports over 
63,700
publications for the search term \texttt{"microservices"}. 
According to Fowler and Lewis~\cite{Fowler2014}, MSA is a service-based architectural style with characteristics like \textit{componentization via services}, \textit{organization around business capabilities}, \textit{infrastructure automation}, and \textit{evolutionary design}.
Creating and operating MS-based systems can be challenging and expensive, and many companies even abandoned MSA for a monolithic architecture~\cite{Noonan2018,Mendonca2021}.
However, a well-designed MSA is beneficial for many software quality attributes, such as maintainability, scalability, reliability, and portability~\cite{Bogner2019,Soldani2018_PainsGainsMicroservices,Wang2021}.

\label{sec:background-devops}
Teams developing MSA-based systems usually follow a 
DevOps life-cycle 
to facilitate the management of a large number of small services~\cite{Chen2018}.
%

\begin{wrapfigure}{r}{0.4\textwidth} 
    \vspace{-0.5cm}
    \centering
    \includegraphics[width=0.4\columnwidth]{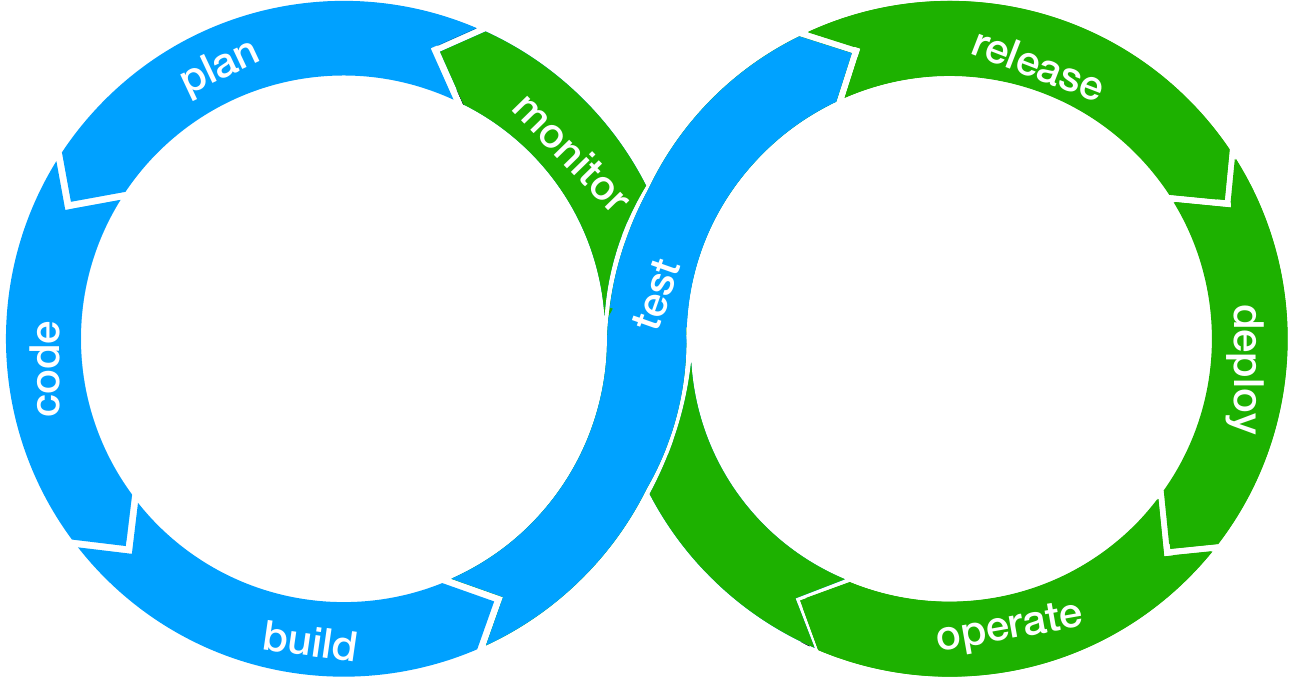}
    \caption{DevOps life-cycle}
    \label{fig:devops-lifecycle}
    \vspace{-0.85cm}
\end{wrapfigure}

DevOps is a software development paradigm trying to bring software faster and more reliably into production~\cite{Bass2015} by destroying barriers between the development and operation teams. 
Another 
objective of DevOps is to reduce cycle time~\cite{Callanan2016}, i.e., how long it takes from 
starting the development of a feature until its deployment in production.
The software development life-cycle is composed of eight phases~\cite{Bass2015}, with the last phase leading back into the first one (see also \autoref{fig:devops-lifecycle}): \textit{Plan}, \textit{Code}, \textit{Build}, \textit{Test}, \textit{Release}, \textit{Deploy}, \textit{Operate}, and \textit{Monitor}. While MSAs can be developed following other life-cycles as well, we will use the DevOps life-cycle as a reference, given that most other development models include a subset of its phases. 
Lately, Machine Learning Operations (MLOps) has been proposed as a version of DevOps enhanced for ML-based applications. Multiple definitions of MLOps have been provided based on the different perspectives of the ML lifecycle. \cite{MLOpsSergio} provides a graphical representation of MLOps aimed at \enquote{maintaining the simple and iconic pipeline of DevOps, yet improving it by adding new circular steps for ML incorporation}. Hence, we could also have used such a MLOps life-cycle representation~\cite{treveil2020mlops} for our analysis. However, it would not have been possible to apply it to other AI techniques; hence, we preferred DevOps which can be applied to a larger set of works. The same applies for the concept of Artificial Intelligence
for IT Operations (AIOps), which makes use of AI for automating the Operational side of DevOps pipelines.

\subsection{Related Work}
\label{sec:rw}
Table~\ref{tab:rw} presents an overview of existing surveys on the use of AI in application architectures, container technologies, and infrastructures that are based on or support MSs. The found surveys belong to three categories, divided by a double horizontal line in Table~\ref{tab:rw}. The works in the first one focus on a set of AI techniques to solve specific problems in specific DevOps phases, while the second category studies the use of very specific AI models for MSs. Finally, the third category lists papers that focus on failure diagnosis and root-cause analysis (RCA) in MSAs by means of AIOps. 

\begin{table}[t]
    \centering
    
    \adjustbox{max width=\textwidth}{

    \begin{tabular}{@{}p{0.2in}|p{0.6in}|p{1,3in}|p{1,7in}|p{0,55in} |p{0,3in}|p{0.3in}@{}}
    \hline
       \textbf{} & \textbf{Models} & \textbf{DevOps Phase}  & \textbf{Studied Problem} & \textbf{Period} & \textbf{\#SPs} & \textbf{Type} \\
    \hline
    Ours & All & All & No restriction on the scope & 2014-2023 & 269 & SMS \\
    \hline
    \cite{hilali2021MsAdaptationUsingML} & ML DL &  Operate Monitor & Self-adaptation of MSs & 2015-2021
    & 62 & SMS \\
    \hline
    \cite{Zhong2021MLbasedOrchestrationofContainers} & ML RL & Deploy Operate Monitor & Performance efficiency of container orchestration & 2016-2021 & 44 & SLR \\
    \hline
    \cite{Duc2019MLMethodsforReliableResourceProvisioning} & ML & Deploy Operate Monitor & Performance efficiency and Re\-lia\-bility of edge-fog-cloud spectrum & n/a & 70 & SLR \\
    \hline
    \cite{Nayeri2021ApplicationPlacementWithAI} & ML EA & Deploy Operate Monitor & Performance efficiency of application placement (broader than MS) in fog computing  & 2017-2020 & 109 & SLR\\
    \hline
    \hline
    \cite{Nguyen2022GNN} & GNN & Monitor Operate Code & Anomaly detection, re\-source scheduling, monolith decomposition & 2020-2022 & 10 & SLR\\
    \hline
    \cite{saucedo2024Clustering} & Clustering  & Plan Code & Monolith decomposition & 2015-2023 & 22 & SMS \\
    \hline
    \hline
    \cite{AIOps_RCA} & AIOps & Operate Monitor & Reliability & n/a & n/a & survey\\
    \hline
    \cite{AIOps_FailureDiagnosis} & AIOps & Operate Monitor & Reliability & 2003-2024 & 94 & SLR\\
    \hline
    \cite{AIOps_Salesforce} & AIOps & Operate Monitor & Reliability & n/a & n/a & survey\\
    \hline
    
    \end{tabular}
    }
    \caption{Comparison of our SMS with the existing surveys. Acronyms: Deep Learning (DL), Reinforcement Learning (RL), Evolutionary Algorithms (EA), Graph Neural Networks (GNN), Genetic Algorithms (GA). Selected Papers (SPs)}
    \label{tab:rw}
\end{table}

Most of the surveys in the first category focus on the use of AI to cost-effectively optimize performance by means of improved support for application placement, autoscaling, and monitoring in the later DevOps phases. Hilali et al.~\cite{hilali2021MsAdaptationUsingML} present an SMS on the use of Machine Learning (ML) for the self-adaptation of MSs. 
Interestingly, the used methodology involves searching papers irrespective of whether they use ML.  An interesting finding is that only 40.3\% of the collected papers on self-adaptation for MSs use either classical ML or Deep Learning (DL) . 
Zhong et al.~\cite{Zhong2021MLbasedOrchestrationofContainers} present a taxonomy and future research directions for ML-based container orchestration via Reinforcement Learning (RL) with a focus on achieving improved performance efficiency.
Duc et al.~\cite{Duc2019MLMethodsforReliableResourceProvisioning} present a survey of ML-based performance modeling and resource management techniques for 
distributed computing environments formed by the spectrum of edge, fog, and cloud computing.
Nayeri et al.~\cite{Nayeri2021ApplicationPlacementWithAI} present a taxonomy of AI-based application placement algorithms for optimizing performance metrics in fog computing environments. 
These algorithms are divided into three groups: Evolutionary Algorithms (EA), ML-based algorithms, and hybrid algorithms that combine different kinds of algorithms. 

The second category of papers focuses on a particular AI technique. It includes two surveys.
Nguyen et al.~\cite{Nguyen2022GNN} present a survey on the use of Graph Neural Networks (GNN) in the field of MSs. 
Saucedo et al.~\cite{saucedo2024Clustering} conducted a systematic mapping study on the use of AI for migrating monolithic application to MS-based applications. A striking finding is that a massive amount of papers have used clustering as the primary technique. In our survey, we also cover GNN papers, which are classified under the neural network keyword that belongs to the ML sub-class. For migration, we also found that clustering and unsupervised learning in general is an often used AI technique.

In the third category of improved reliability by means of AIOps, we have found 3 preprints on arXiv. The oldest work by Salesforce~\cite{AIOps_Salesforce} studies AIOps for cloud computing with a partial focus on MSs, where it reviews the use of AI for incident detection, failure prediction and RCA. Moreover, it defines three different data sources for AIOps in cloud computing: (1) metric-based data, (2) heterogeneous log data and (3) traces, uncovering not only the dynamic topological structure of MSs but also generating multi-modal data by combining it with the two previous data sources. The two other surveys~\cite{AIOps_FailureDiagnosis},~\cite{AIOps_RCA} focus  respectively on failure diagnosis and RCA of MSs. Similarly to \cite{AIOps_Salesforce}, they consider metrics, logs, or traces as relevant data sources, but they also consider 
multi-modal approaches that combine metrics and logs~\cite{AIOps_FailureDiagnosis}. 
Our survey pointed to 50 papers that use AI for improving reliability of MSs during the Ops stage (i.e., the "Deploy", "Operate", "Monitor" phases) and also identified better support for AIOps in MSs as a new trend (cfr. Section \ref{sec:sa_qa_ai}). 

Unlike these surveys, we aim to understand the complete panorama of the use of AI in all DevOps phases, without restricting the survey to a specific problem. 
Thus our survey covers many more primary studies than the existing surveys. As a side effect, we need to remain at a more abstract level in the analysis. 


\section{Methodology}
\label{Methodology}

Our systematic mapping study is based on the guidelines defined by Petersen et al.~\cite{PETERSEN20151}. We also applied the \enquote{snowballing} process defined by Wohlin~\cite{Wohlin2014}. In this section, we describe the goal and the research questions (Section~\ref{RQ}), report our search strategy approach 
and outline the data extraction and the analysis of the corresponding data (Section~\ref{Strategy}). 
The list of selected papers (SPs) is provided as a supplementary material\footnote{\url{https://doi.org/10.6084/m9.figshare.26243993}} due to space constraints.


\subsection{Goal and Research Questions}
\label{RQ}
As anticipated in the introduction, our goal is to analyze the use of AI techniques to solve the challenges posed by the design, development, and operation of MS systems.
Based on it, we first conducted a preliminary study aiming to analyze the trend of the research on AI4MS by performing historical analysis. Specifically, we assessed how the number of AI4MS publications has evolved over the years. Then, we designed the remainder of our study around the following research questions (\textbf{RQs}):

\begin{tabular}
{@{}p{0.4cm}p{11.5cm}@{}}
\textbf{RQ$_1$} & 
In which industry domains is AI used for MSs?\\
\textbf{RQ$_{2}$} & 
Which quality attributes are improved by AI4MS?\\
\textbf{RQ$_{3}$} & 
In which DevOps phases is AI4MS applied?\\
\textbf{RQ$_4$} & 
What AI techniques are used for realizing AI4MS?\\
\textbf{RQ$_5$} & 
What are the open challenges in AI4MS?\\
\end{tabular}
\vspace{2mm}


The defined research questions focus on the technical and scientific contents of the research. First, \textbf{RQ$_1$} analyses which industry domain the approach targets. We then want to capture why and when AI has been applied, and the answer is typically in terms of improving some quality attribute in the context of some specific phase of the DevOps development life-cycle. Notably, even if the DevOps life-cycle is very common in MSs, we do not intend to disregard approaches based on different life-cycles. Still, they can normally be mapped into subsets of the phases of the DevOps cycle.
\textbf{RQ$_{2}$} focuses on the improved quality attributes, while \textbf{RQ$_{3}$} discusses the DevOps phase where such improvement occurs. 
With \textbf{RQ$_4$} we want to investigate what AI techniques have been used in the selected works.
The answers to RQs 2-4 will first be discussed separately, and then they will be combined by means of \emph{a multidimensional analysis} that identifies interesting connections between AI techniques (RQ4) (and the rationale for using them) and the combination of quality attributes (RQ2) and DevOps phases (RQ3), i.e., what AI technique is often applied to what quality attributes during which DevOps phases and why.
Finally, \textbf{RQ$_5$} highlights the challenges that need to be tackled in future research in the area. 

\subsection{Search Strategy}
\label{Strategy}
\begin{figure}[h]
\centering
\includegraphics[width=0.8\linewidth,clip]{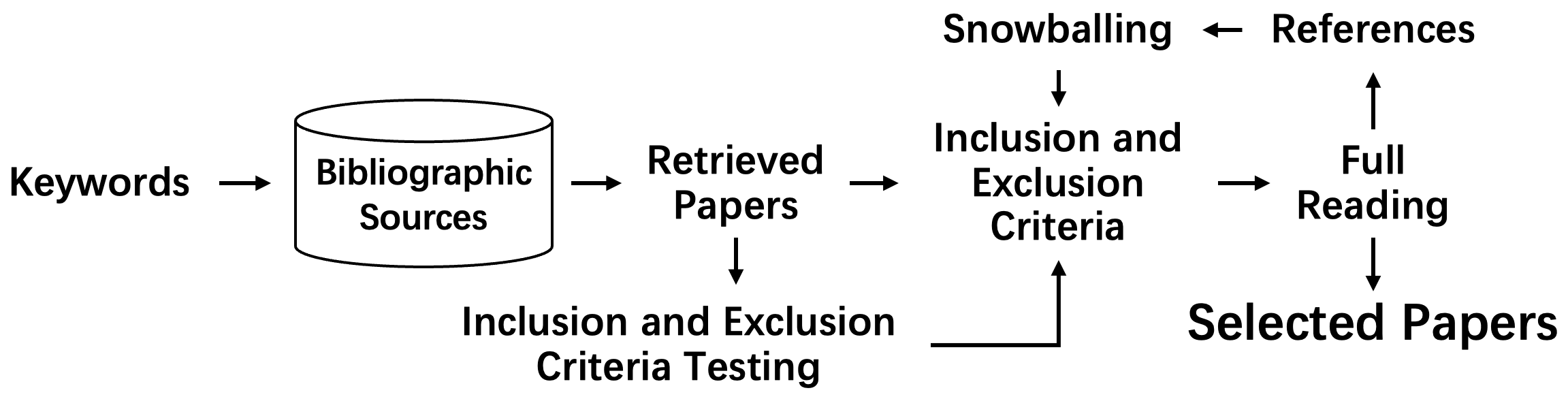}
\caption{The search and selection process}
\vspace{-0.5cm}
\label{fig:SelectionProcess}
\end{figure}

The search strategy involves the outline of the most relevant bibliographic sources and search terms, the definition of the inclusion and exclusion criteria, and the selection process relevant to the inclusion decision. Our search strategy is depicted in Figure~\ref{fig:SelectionProcess}.

\textbf{Search terms}. Our search string consists of a bucket of different microservices spellings and a bucket of various AI-related keywords. We arrived at this search string by prototyping several queries and then iteratively refining the most promising candidate. We aimed for a broad coverage, while simultaneously trying to keep the number of false positives low. The concrete search string looks as follows: 

\vspace{2mm}
\begin{centering}
\textit{("microservic*" OR "micro-servic*" OR "micro servic*") AND ("AI" OR "artificial intelligence" OR "machine learning" OR "machine-learning" OR "ML" OR "deep learning" OR "deep-learning" OR "neural" OR "intelligen* learning*")} 
\end{centering}

\vspace{2mm}

\textbf{Bibliographic sources}. We selected the list of relevant bibliographic sources following the suggestions of Kitchenham and Charters~\cite{Kitchenham2007} since these sources are recognized as the most representative in the software engineering domain and used in many secondary studies. The list includes: \textit{ACM Digital Library, IEEEXplore Digital Library, Scopus, Google Scholar, Springer link}. 

\textbf{Inclusion and exclusion criteria}. We defined inclusion and exclusion criteria to be applied to the bibliographic information (B), to title and abstract (T/A), or to the full text (F), or to both the two last items (Both), as reported in Table~\ref{tab:Criteria}.
A main point is that the search, being keyword-based, naturally resulted in extracting both papers about applications of AI to MSs, relevant for our survey, and papers studying the use of MSs for supporting AI (mostly about using MS systems to support the execution of AI applications). We wanted to focus on the first class of papers; hence, the second class was discarded by our inclusion and exclusion criteria.

\vspace{-0.5cm}
\begin{table}[!ht]
\vspace{-0.5cm}
\centering
\footnotesize
\caption{Inclusion and exclusion criteria} 
\label{tab:Criteria} 
\begin{tabular}
{@{}p{1.4cm}|p{7.5cm}|p{0.9cm}@{}}
\hline
\textbf{Criteria} &\textbf{ Assessment Criteria} & \textbf{Step} \\ \hline
Inclusion & Papers discussing applications of AI to MSs & Both 
\\ \hline
\multirow{8}{*}{Exclusion} & Not fully written in English  & T/A \\
& Non peer-reviewed & B \\
& Books & T/A\\ 
& Duplicated & T/A\\ 
& Full text inaccessible to us & F\\ 
& Out of topic & Both\\ 
& Published before \cite{Fowler2014} (i.e.~older than 2014) & B \\ 
\hline
\end{tabular}
\vspace{-0.7cm}
\end{table}
\begin{table}[h]
\vspace{-0.5cm}
\centering
\footnotesize
\caption{Results of search and selection} 
\label{tab:SelectionResults} 
\begin{tabular}
{|@{}p{6.3cm}|p{2cm}@{}|}  
\hline
\textbf{Step} & \multicolumn{1}{|c|}{\textbf{\# Papers}}  \\ \hline
Retrieval from bibliographic sources (unique) & 3991\\ \hline 
Reading by title and abstract (rejected)  & 3177 \\ \hline 
Full reading (rejected) & 614\\ \hline 
Backward and forward snowballing (accepted) & 69\\ \hline 
\textbf{Primary Studies} & \textbf{269} \\ \hline 
\end{tabular}
\vspace{-0.2cm}
\end{table}

\textbf{Search and selection process}. The search was conducted in September 2023 and included all the publications available until then. The application of the search terms returned 3,991 unique papers.


\textit{Testing the applicability of inclusion and exclusion criteria:} Before applying the inclusion and exclusion criteria, we tested their applicability~\cite{Kitchenham2013} on a subset of 50 retrieved papers (each assigned to two authors), randomly selected.

\textit{Applying inclusion and exclusion criteria to bibliographic information, title, and abstract:} We applied the refined criteria to the remaining 3,941 papers. 
Two authors read each paper; in case of disagreement, at least one additional author was involved in the discussion to clear up any such disagreement.
For 142 papers, we involved more than two authors. Out of the 3,991 initial papers, we included 814 papers based on titles and abstracts.
We adopted \textit{adaptive reading depth}~\cite{Petersen2008} for initial inclusion: in case it was unclear from the title and abstract whether the paper was about the use of AI for MSs, we skimmed through the main text to get a more informed opinion. To measure the level of agreement among the authors at this stage, we computed Cohen's Kappa coefficient~\cite{Kappa}, which resulted in an almost perfect agreement (\textbf{0.889}).

\textit{Full reading:}
We fully read the 814 papers included by title and abstract, applying the same criteria defined in Table~\ref{tab:Criteria} and assigning each one to two authors. We involved a third author for 55 papers 
to reach a final decision. Based on this step, we selected 200 papers 
as relevant contributions. The application of the inclusion and exclusion criteria resulted in an almost perfect agreement (Cohen's Kappa coefficient = 
\textbf{0.839})~\cite{Kappa}.

\textit{Snowballing:} We performed the snowballing process~\cite{Wohlin2014}, considering all the references presented in the retrieved papers and evaluating all the papers referencing the retrieved ones. We applied the same process as for the retrieved papers. The snowballing search was conducted in January 2024, considering all papers published up to 2023 (papers after September 2023 were indeed only retrived by snowballing).
We identified 158 potential papers 
but only 69 of these were included to compose the final set of publications. 

Based on the search and selection process, we retrieved a total of 269 papers for the review, as reported in Table~\ref{tab:SelectionResults}.

\textit{Quality Assessment:}
\label{QualityAssessment}
We decided not to perform any further quality assessment, as this is common for systematic mapping studies that want to provide an overview of the research landscape. The only quality control happened through the focus on peer-reviewed publications. Since AI4MS is a very young field, many approaches are also still preliminary, and a too-strict quality assessment may remove papers that are a first attempt towards a promising approach.


\textbf{Data Extraction and Replicability:}
\label{DataExtraction}
We extracted data from the selected Primary Studies (PSs). 
The data extraction form, together with the mapping of the information needed to answer each RQ, is summarized in Table~\ref{tab:DataExtraction}.


\begin{table}[]
\adjustbox{max width=\textwidth}{
    \centering
    \begin{tabular}
{l|l|p{6cm}}%
\hline
\textbf{RQ} & \textbf{Info} & \textbf{Description}  \\ \hline
All & Title, Authors, DOI, Abstract, Publication Venue & Main information 
\\ \hline
Preliminary analysis & Year &  \\  \hline
RQ1 & Industry Domain & Domain in which the work has been applied (divided in Level 1 and Level 2)  \\ \hline
RQ2.1; RQ2.3 & Improved quality attributes & According to ISO25010\\ 
\hline
RQ3; RQ2.3 & Improved DevOps phases & \\ \hline
RQ4 & AI Model & According to the taxonomy in~\cite{AIWatch}\\  \hline
RQ5 & Future challenges & Future work and challenges\\
\hline
\end{tabular}
}
\caption{Data extraction} 
\label{tab:DataExtraction}
\end{table}

To allow one to trace the data extraction process, we prepared a replication package\footnote{\url{https://doi.org/10.6084/m9.figshare.22663756} \label{Package}} for this study with the complete results obtained.
This would also allow replication and extension of our work by other researchers.

\section{Results}
\label{Results}



\subsection{Preliminary analysis on the number of AI4MS publications}

\begin{wrapfigure}{r}{0.5\textwidth}
    \centering
    \includegraphics[width=0.5\columnwidth]{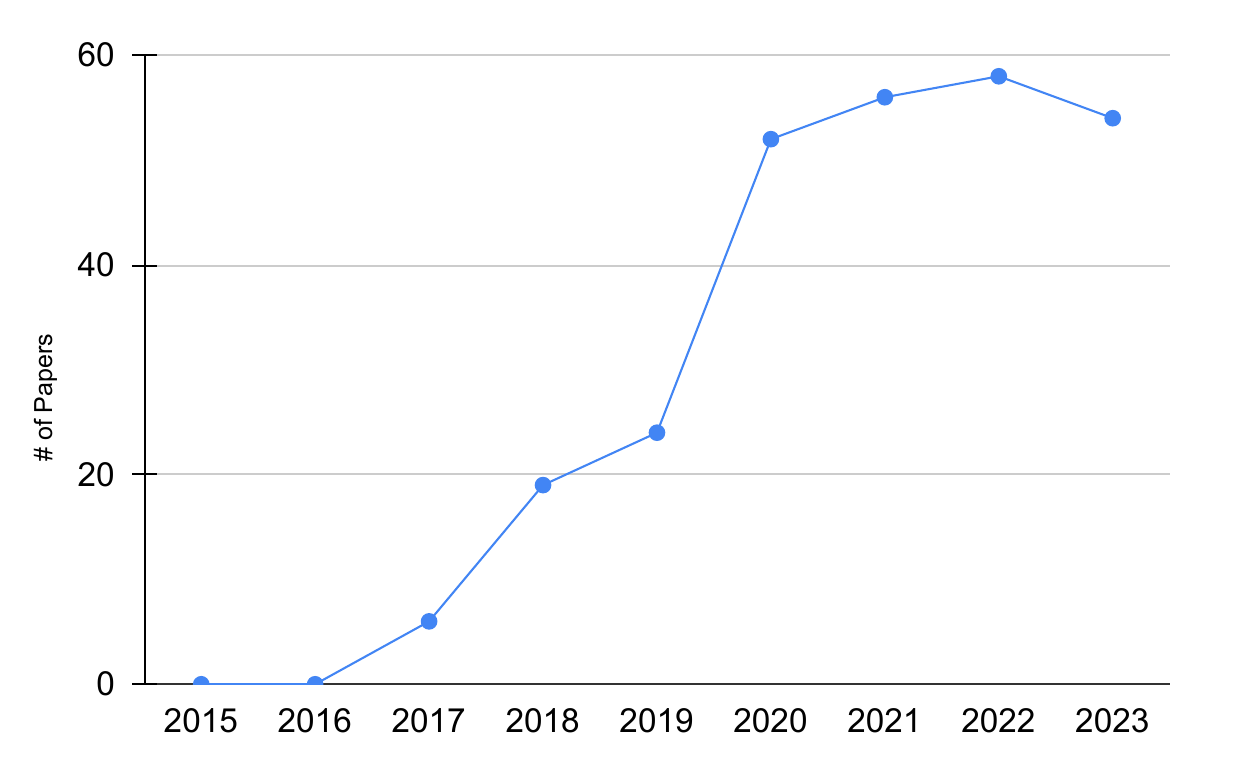}
    
    \caption{Number of publications per year}
    \vspace{-0.6cm}
    \label{fig:publications-year}
\end{wrapfigure}

\autoref{fig:publications-year} shows the evolution of AI4MS publications from 2016 to 2023.
We can observe that it took three years from the 2014 blog post by Martin Fowler and James Lewis~\cite{Fowler2014} to start considering the use of AI techniques to support MSs in their development, deployment, and runtime management.
Indeed, the first three years of MS-related research were mainly devoted to understanding the advantages, drawbacks, and potentials of MSs, as outlined, e.g., in \cite{Soldani2018_PainsGainsMicroservices}.

In 2017, MSs were already widespread, with big IT players (e.g., Amazon, Netflix, and Spotify) using them to deliver their core businesses \ref{SP114}.
This raised interest in how to better support MSs, and researchers started using AI to realize such support. Since 2017, we indeed have had an ever-increasing trend of AI4MS publications, witnessing a wider and wider recognition of the potential of AI to support MSs.

However, starting in 2020, the trend of sharp increases in AI4MS slowed down, and indeed the numbers of publications through 2020 to 2023 are nearly identical (the small decrease in the last year should be considered with care, since it may be partially due to delays in publication and indexing of some papers). 

\subsection{RQ1. In which industry domains is AI used for MSs?}
To classify the selected studies based on targeted industry domains, we started from the taxonomy of economic sectors defined by AIWatch \cite{AIWatch}. 
The latter enables distinguishing the application of AI to different industry domains. 
Unsurprisingly, being MSs themselves part of the \textit{information and communication} industry, 255 of the selected studies pertain to such an industry domain, with an ever-increasing trend since 2017 (in line with the results discussed in our preliminary analysis).
We also observed a recent interest in using AI for MSs in the \textit{manufacturing} field, with \ref{SP114},\ref{SP116},\ref{SP138} showing that AI is now starting to get used to support MSs in realizing cyber-physical systems for the Industry 4.0 paradigm. 

Given that the vast majority of the selected studies pertained to the \textit{information and communication} industry domain, we mapped them to well-known sub-domains.
%


%

\begin{figure}
\vspace{-0.5cm}
\centering
  \includegraphics[width=0.7\linewidth]{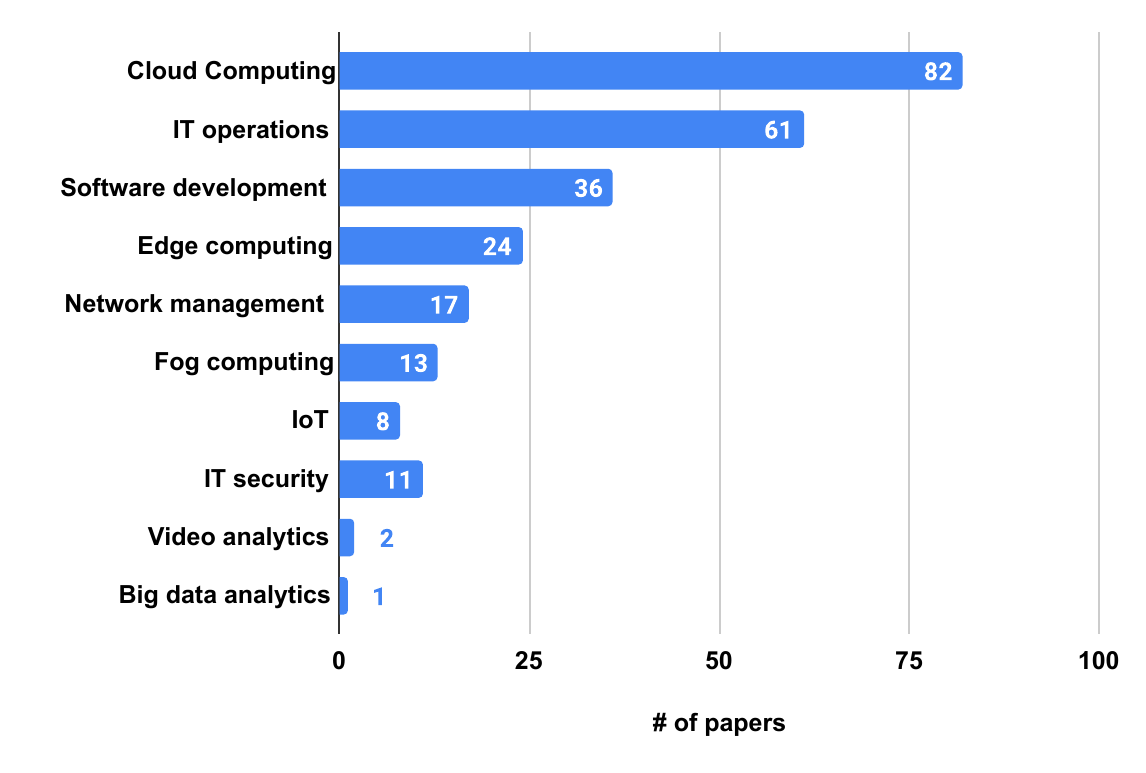}
  \captionof{figure}{Sub-domains of \textit{information and communication} where AI is used for MSs.}
  \label{fig:industry-domain-level2}
\end{figure}
\begin{figure}
  \centering
  \includegraphics[width=.7\linewidth]{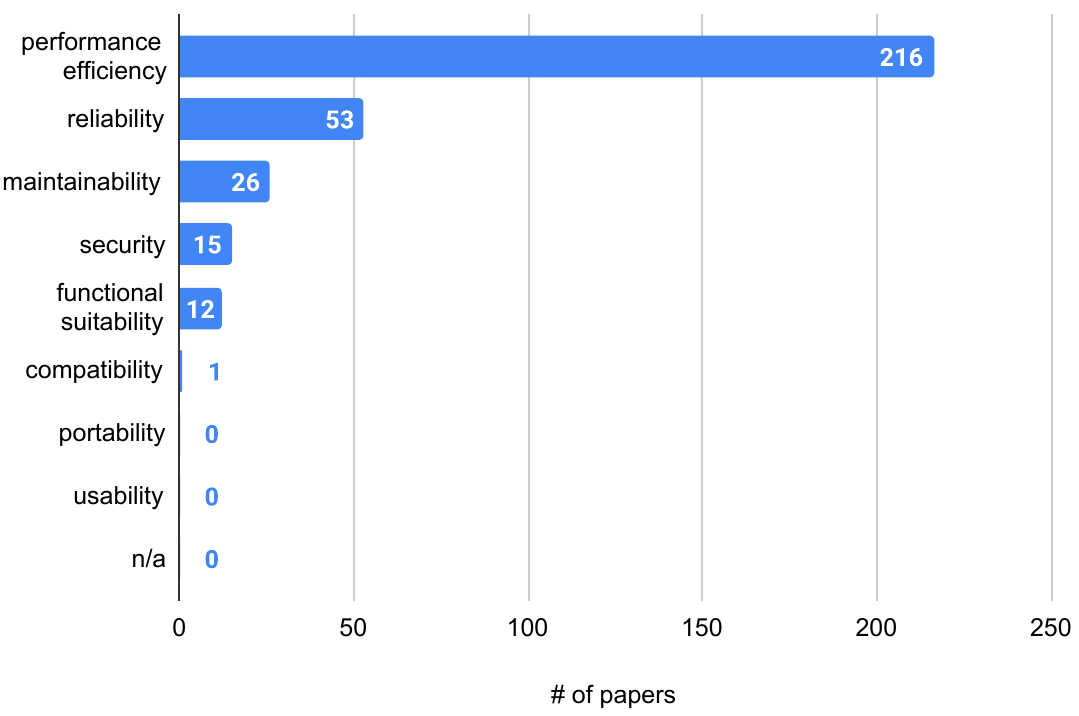}
  \captionof{figure}{Microservices quality attributes improved through the use of AI}
  \label{fig:QAs}
\vspace{-0.5cm}
\end{figure}

The result is shown in \autoref{fig:industry-domain-level2}, from which we observe that AI is mostly used to support MSs in \textit{cloud computing}.
This is somewhat expected, as one main advantage of MSs is to enable realizing cloud-native applications \cite{Soldani2018_PainsGainsMicroservices}, which makes \textit{cloud computing} their natural industry sub-domain. 

The significant coverage of \textit{edge computing} and \textit{fog computing} aligns with the above considerations.
Indeed, edge and fog computing are intended to enable computations to happen closer to IoT, either fully there or by creating a sort of computing continuum from cloud to IoT.
This is done by distributing the services forming an application on computing devices that are physically close to Things, also exploiting virtualization, and similarly to what happens in-cloud; however, considering the locality of the computation and the fact that such devices have limited computing resources. 
For instance, 24 of the selected primary studies illustrate how MSs can be exploited to realize edge applications, as shown in \autoref{fig:industry-domain-level2}.
The figure points out that AI can support MSs in \textit{edge computing} and \textit{fog computing}, e.g., for resource provisioning~\ref{SP35}, \ref{SP61}, MSs' scheduling~\ref{SP32}, \ref{SP50}, or their runtime management~\ref{SP1}, \ref{SP39}.

Another insight follows from the significant coverage of DevOps among selected studies pertaining to the \textit{information and communication} industry domain. 
Indeed, \textit{software development} and \textit{IT operations} are targeted by 36 and 61 selected studies, respectively.
On the \textit{software development} side, AI is mostly used to automate the migration of existing applications to MSs, e.g., \ref{SP31}, \ref{SP76}, \ref{SP94}, \ref{SP107}.
On the \textit{IT operations} side, AI is instead used for multiple tasks, e.g., auto-scaling \ref{SP125}, \ref{SP131} or fault diagnosis \ref{SP126}, \ref{SP134}. 
This showcases the potential of AI to support the DevOps activities for MSs, with an increasing trend since 2017, making this a promising research direction.
Finally, in \textit{IT security}, covered by 11 studies,
AI is used to automate intrusion detection, typically based on detecting anomalously behaving MSs, e.g., \ref{SP49}, \ref{SP69}, \ref{SP81}, \ref{SP137}. 
Despite low numbers, it started being considered only in 2018, with an overall increasing trend since then due to promising results.
The use of AI for MSs in the domain of \textit{IT security} hence deserves further investigation.

\subsection{RQ2. Which quality attributes are improved by AI4MS?}\label{sec:qa}
To classify the papers according to the improved quality characteristics, we used the well-known ISO 25010:2011 standard \enquote{Systems and software Quality Requirements and Evaluation} (SQuaRE)~\cite{InternationalOrganizationForStandardization2011}.
It contains a software product quality model with eight different top-level quality attributes (QAs), i.e., functional suitability, performance efficiency, compatibility, usability, reliability, security, maintainability, and portability. 
An attribute was assigned to a paper if the described use of AI was intended to improve this QA.
No sub-QAs, e.g., time behavior or capacity for performance efficiency, or QAs outside of ISO 25010:2011, e.g., scalability or observability, were used.




Each of the 269 papers was assigned either 1 or 2 improved QAs, with no paper using AI to simultaneously improve 3 or more QAs. For two QAs, namely portability and usability, we did not find any papers. The distribution of the eight QAs is shown in \autoref{fig:QAs}. As visible, the sample is dominated by \textit{performance efficiency} (216 papers, $\sim$80\%). For most of these papers, the goal was to improve the scalability of MS-based systems, i.e., to increase throughput while simultaneously keeping response times small. AI-powered approaches to achieve this were, e.g., service auto-scaling techniques~\ref{SP6}, \ref{SP17}, \ref{SP55}, sophisticated load-balancing~\ref{SP7}, \ref{SP35}, \ref{SP139}, or dynamic service placement within a cloud-fog-edge-continuum~\ref{SP18}, \ref{SP20}, \ref{SP33}.
Most of these 216 papers focused exclusively on performance efficiency (171, $\sim$80\%).
However, several papers also combined this QA with \textit{reliability}: 34 of the 53 papers with reliability also improved performance efficiency (64\%).
Such papers either explicitly added availability as a targeted QA for their auto-scaling \ref{SP16}, scheduling \ref{SP27}, or load-balancing \ref{SP36} approach or used AI to reduce service downtime by identifying anomalies and faults~\ref{SP23}, \ref{SP126}, \ref{SP135}, \ref{SP233}, \ref{SP134}, \ref{SP238}.

Other quality attributes were less prominent in our sample.
A total of 26 papers used AI to improve \textit{maintainability}.
These were usually approaches to help with architecting MSs, e.g., by using AI to propose how to decompose a monolithic application into microservices~\ref{SP136}, \ref{SP247}, \ref{SP272}, suggesting detailed migration plans~\ref{SP31}, or AI-powered approaches for architectural runtime adaptation~\ref{SP93}, \ref{SP96}.
Maintainability papers were sometimes paired with performance efficiency, functional suitability, or reliability, but 11 papers also focused exclusively on maintainability.
Similarly, 15 papers improved \textit{security}, and all but 5 of these papers did so exclusively.
Most of these approaches used AI to identify security-relevant anomalies and malicious behavior, e.g., by analyzing service communication traces~\ref{SP21}, \ref{SP81}, \ref{SP137}.
Furthermore, 12 papers improved \textit{functional suitability}.
These were usually AI-based approaches for automatically recommending suitable services for composition~\ref{SP53}, \ref{SP76}, \ref{SP122}.
Some papers also used natural language models as BERT~\ref{SP53} or GPT~\ref{SP235}  to analyze natural language requirements and to propose suitable microservices based on them.

Lastly, a single paper used AI to improve compatibility~\ref{SP269}: in the context of wireless sensor networks, the authors propose a deep learning approach for microservice interoperability to allow dynamic service interactions. AI-based approaches to improve the \textit{portability} or \textit{usability} of microservices did not appear in our sample.


\subsection{RQ3. In which DevOps phases is AI4MS applied?}
\label{sec:devops}
\begin{figure}[!ht]
\vspace{-0.3cm}
\centering
  \includegraphics[width=0.8\linewidth]{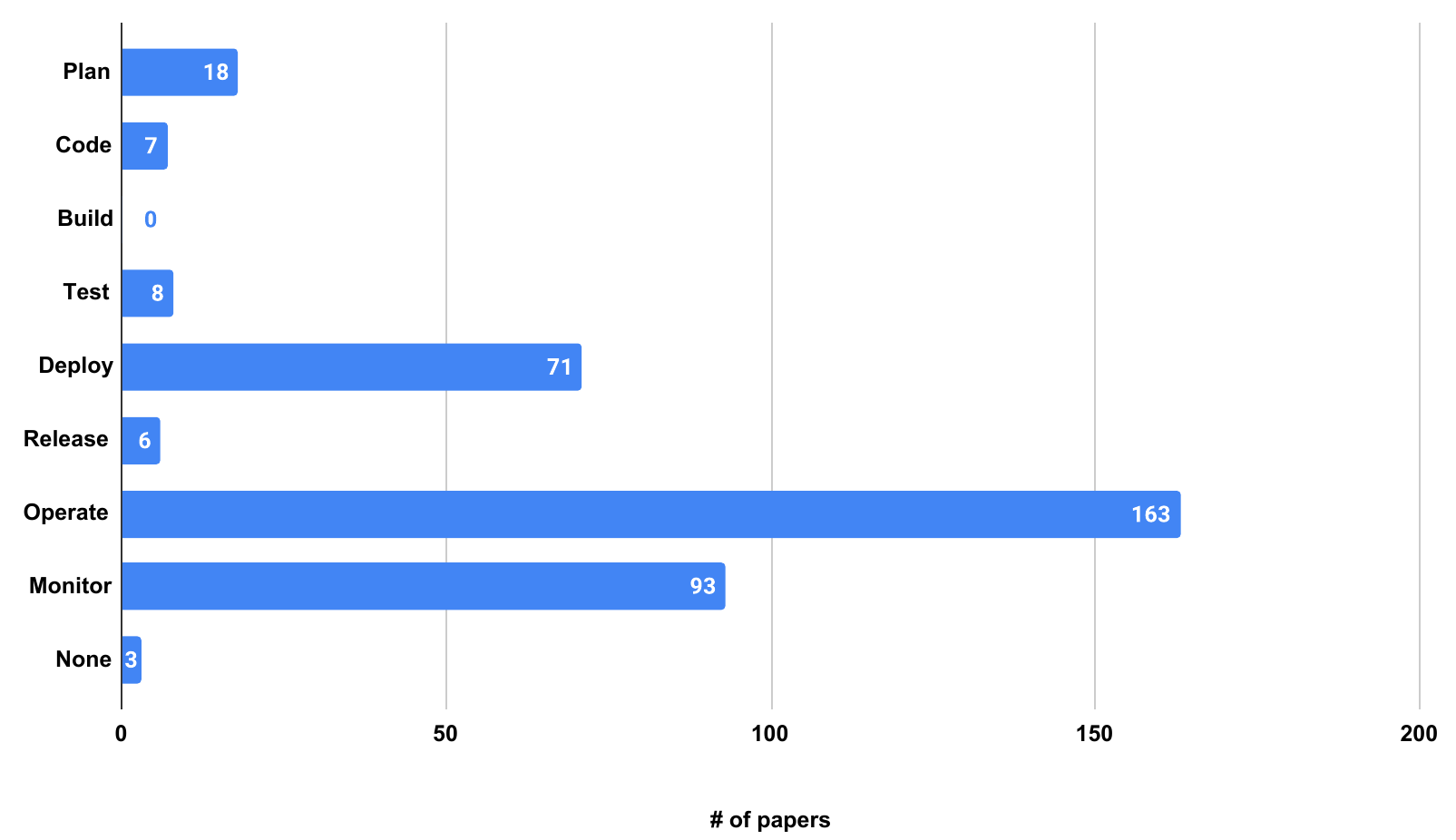}
  \captionof{figure}{DevOps phases improved by using AI}
  \label{fig:devops}
\end{figure}

To classify the papers according to the improved software engineering activities, we used the well-known DevOps life-cycle phases presented in Section~\ref{sec:background-devops}.
The analysis performed is depicted in \autoref{fig:devops}. We can see that of the 269 papers, only 3 could not be traced to specific DevOps phases, while the rest improved at least a single phase. Unsurprisingly, we can see that none of the works contribute to the \textit{Build} phase, as code compilation does not need AI.



Most of the papers focuses on the adoption of AI techniques for improving the \textit{Operate} phase of MSs. Following the discussion in Section~\ref{sec:background-devops}, it is not surprising that the rise of AIOps, pushed approaches on using AI to improve the \textit{Operate} and \textit{Monitor} phases~\ref{SP143}~\ref{SP228}. The majority of papers addressing the Operate phase focus on the concept of \textit{Scaling} as a central theme. Such emphasis reflects an increasing necessity to guarantee that systems and solutions can efficiently accommodate augmented workloads, user demands, or data volumes without jeopardising performance or reliability. The main goal of these projects is to use AI technologies to create and improve auto-scaling mechanisms, making the system more intelligent and adaptive. This would ensure the best use of resources while keeping the costs down and avoiding performance issues~\ref{SP16}~\ref{SP28}~\ref{SP36}. Among these techniques, Reinforcement Learning (RL) has gained significant attention in the context of managing scaling in unpredictable and highly variable workloads due to its flexibility and self-improving nature~\ref{SP44}~\ref{SP204}.

\begin{figure}[!ht]
    \centering
    \includegraphics[width=0.8\columnwidth]{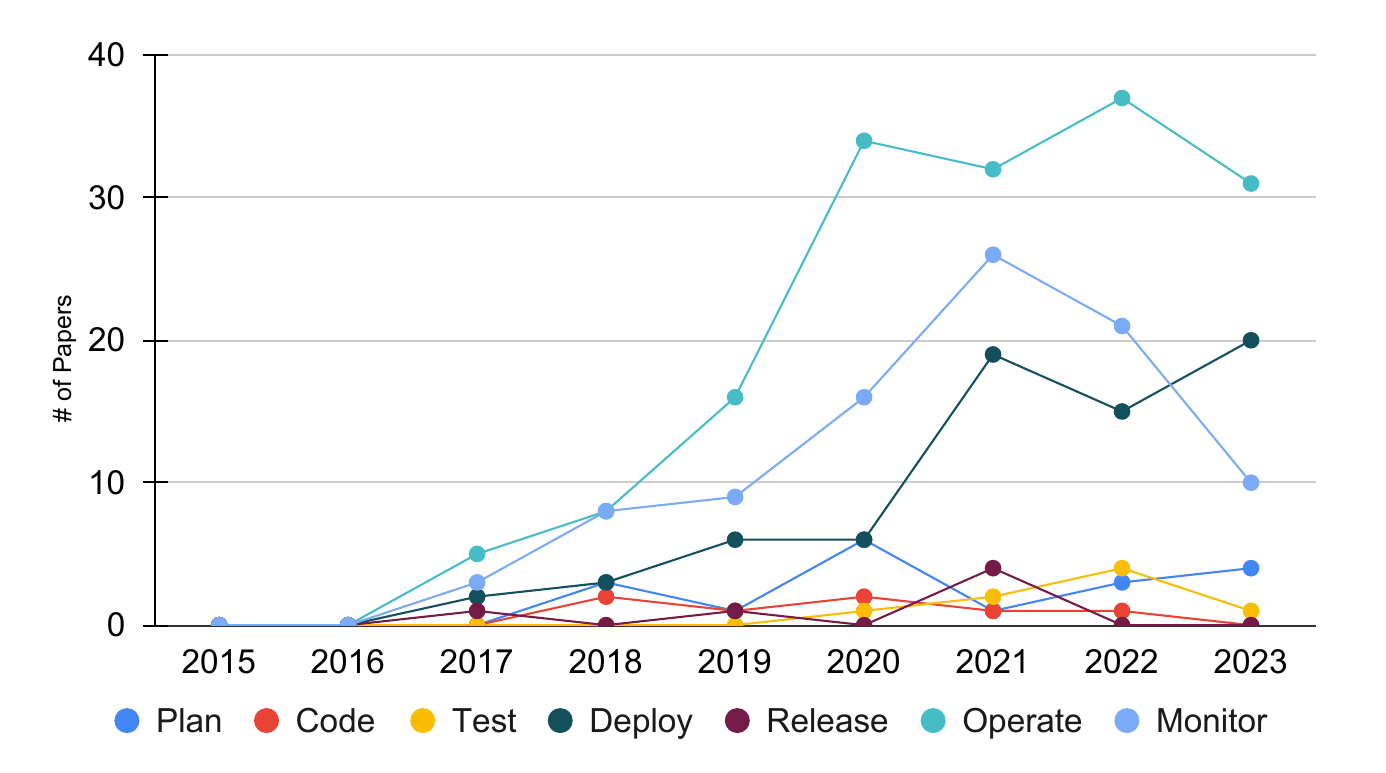}
    \caption{DevOps phases improved by using AI by years}
    \vspace{-0.5cm}
    \label{fig:devopsyear}
\end{figure}

Furthermore, we conducted an analysis on the trend of publications regarding the adoption of AI to improve each DevOps phase (shown in Figure \ref{fig:devopsyear}). Specifically concerning the three phases with the most publications, i.e., \textit{Operate}, \textit{Monitor} and \textit{Deploy}, we can observe that from 2021 the increase of the number of papers on each of these three phases either decelerate or even decrease in numbers, compared to those of the previous years. Nonetheless, the publications concerning these three phases are still more than the ones considering the other phases.


\subsection{RQ4. What AI techniques are used for realizing AI4MS?}
\label{sec:ai}


To understand the AI technique used in the selected works, we classified them according to the taxonomy in~\cite{AIWatch}. We used both AI domains, for a coarse-grained analysis (useful, e.g., to understand the time evolution of the field), and keywords and AI subdomains together, for a more detailed analysis. We used AI subdomains together with keywords (and below, for conciseness, we will refer to both of them only as keywords) since it was not always possible to assign a specific keyword to an approach, e.g., since the work uses a family of related techniques.
We remark that keywords are not orthogonal, and that a single approach may involve multiple keywords. Indeed, we assigned from one to five keywords per paper. 
Also, some keywords are pretty general (e.g., neural networks), hence one could naturally expect higher frequencies. However, when multiple keywords would be compatible with an approach, we preferred specific keywords to more general ones.    

\begin{figure}
    \centering
  \hspace{-2cm}\includegraphics[width=0.8\linewidth]{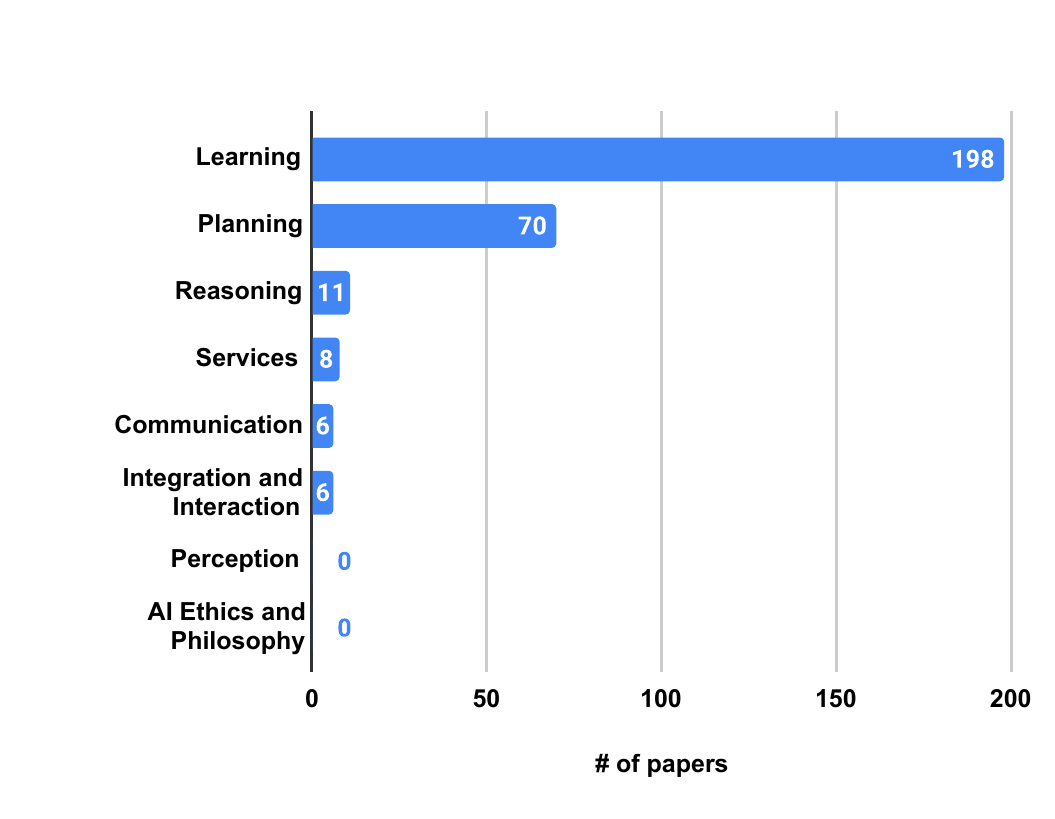}
  \captionof{figure}{AI domains of techniques applied to MSs}
  \label{fig:AIDomains.pdf}
\end{figure}

\begin{figure}[!ht]
    \vspace{-0.5cm}
    \hspace{-.6cm}\includegraphics[width=1.06\columnwidth]{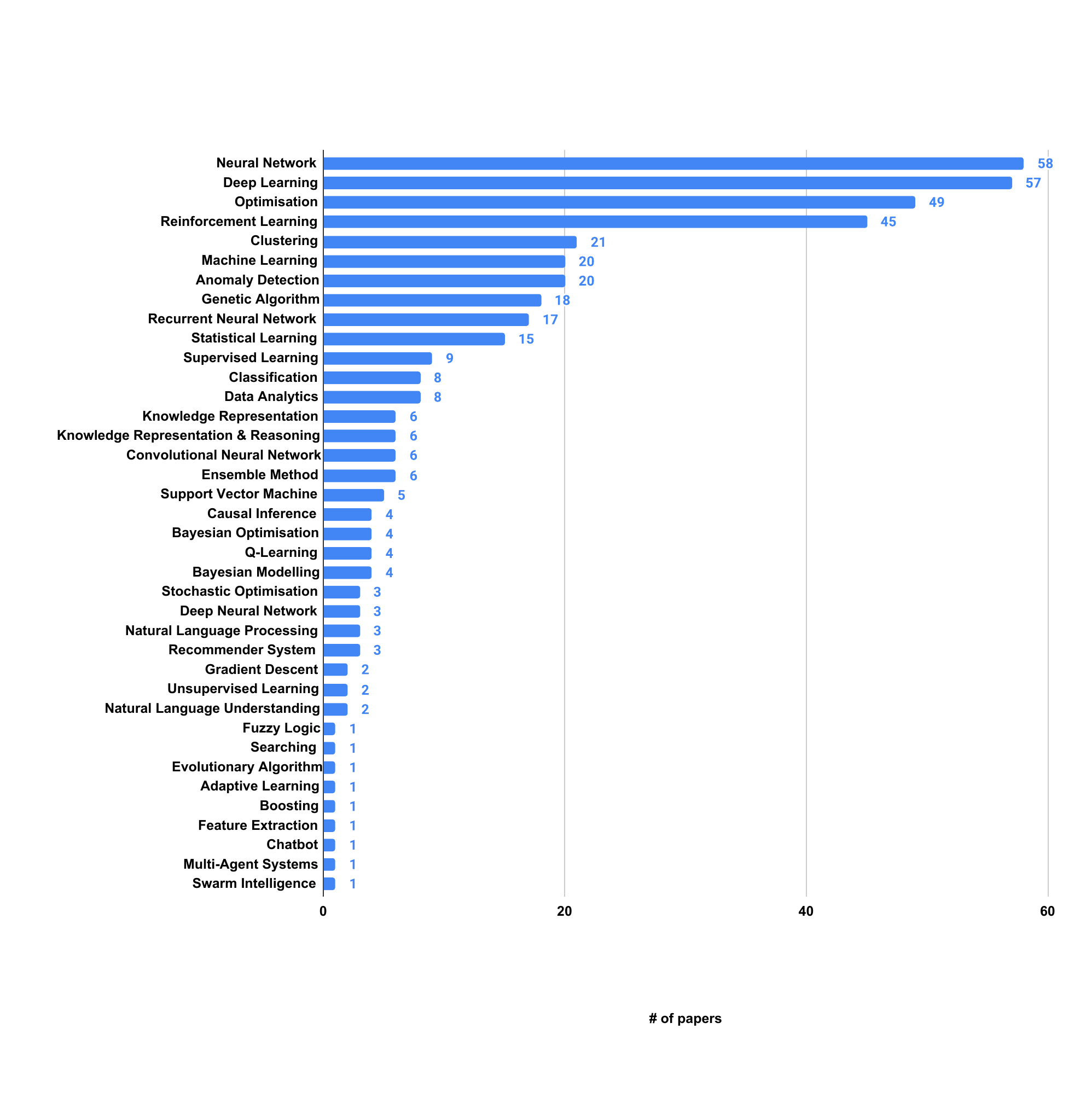}
    \caption{AI keywords related to approaches applied to MSs}
    \vspace{-0.5cm}
    \label{fig:AIkeywords}
\end{figure}

Results are shown in Figure~\ref{fig:AIDomains.pdf} for AI domains and in Figure~\ref{fig:AIkeywords} for keywords. Both the figures just show the frequency of each item. For domains, learning is by far the most frequent, followed by planning and reasoning, with all the other AI domains taking a marginal role. The relevance of learning is confirmed by looking at keywords, with many keywords in the domain having a high frequency. This is expected since such approaches are useful to tackle a number of software engineering problems, and MSs are no exception. Inside the domain, there is no clear winning approach, with various keywords scoring high, and the top places being taken by more general keywords (neural network, deep learning, reinforcement learning, etc.).
Apart from these, the single keyword which scores highest is optimization from the planning domain, which finds obvious applications to find the best configurations to optimize relevant QAs. This is in line with the observation of Section~\ref{sec:qa} that the most considered QA is performance efficiency which, being quantitative, can benefit from optimization. Indeed, a number of works deal with optimization for various aspects of performance efficiency. This is for instance the case of~\ref{SP141} which tackles optimization of task scheduling in mobile Cloud computing, of~\ref{SP172} which considers application placement and migration in the Cloud-IoT continuum, and~\ref{SP266} which deals with deployment and startup of microservice instances in resource centres. However, optimization can also be used for other QAs, e.g., it is used in~\ref{SP159} to find microservice candidates in the refactoring of legacy systems into microservice architectures to optimize maintainability metrics such as feature modularization and reuse.  
Another frequent keyword is anomaly detection (from the learning domain), suitable for highlighting anomalous behaviors that need to be managed. Interestingly, anomalies are mostly related to performance efficiency, frequently paired with reliability~\ref{SP13}, \ref{SP23}, but some of them are also related to security~\ref{SP161}.  Indeed,~\ref{SP13} shows that by optimizing the performance of serverless systems by reducing the number of cold starts of functions, one is also able to reduce the number of failed calls. Instead,~\ref{SP161} looks for anomalies in logs of API invocations to highlight data breaches and DoS attacks.
There are a few works in the domain of communication, which used to be focused on natural language understanding, such as~\ref{SP163} where it is used to extract information from user specifications. However, in 2023 the first work exploiting large language models (chatGPT in the specific case) for microservices~\ref{SP235} appeared. We expect such line of work to get considerable attention in the next years.

An analysis of the evolution of AI domains over the years is actually meaningful only for learning and planning, since the other domains have low frequencies. Learning had a relevant growth and is now essentially stable, hence the approach is probably reaching maturity. A similar trend is also visible for planning, albeit the growth ended up earlier. 
A deeper analysis of the involved works reveals that planning is only applied during Ops phases, while learning is applied to both Dev and Ops phases, therefore suggesting better compatibility of ML with the entire DevOps lifecycle.

\subsection{RQ5. What are the open challenges in AI4MS?}
\begin{figure}[!t]
\vspace{-0.5cm}
    \centering
    \hspace{-2cm}\includegraphics[width=1\columnwidth]{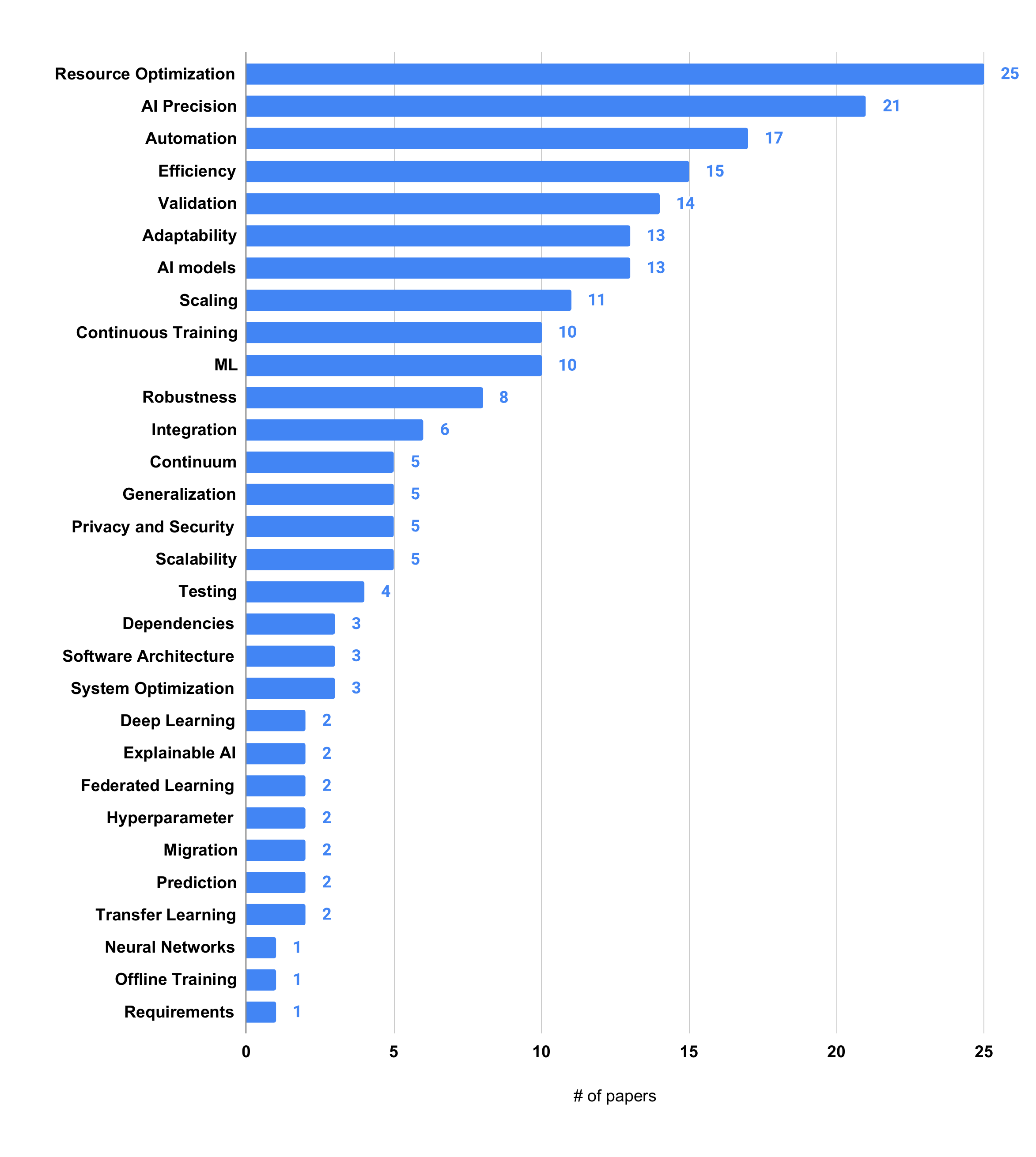}
    \caption{Future challenges}
    \vspace{-0.5cm}
    \label{fig:fut}
\end{figure}

\begin{figure}[!t]
    \centering
    \includegraphics[width=1\columnwidth]{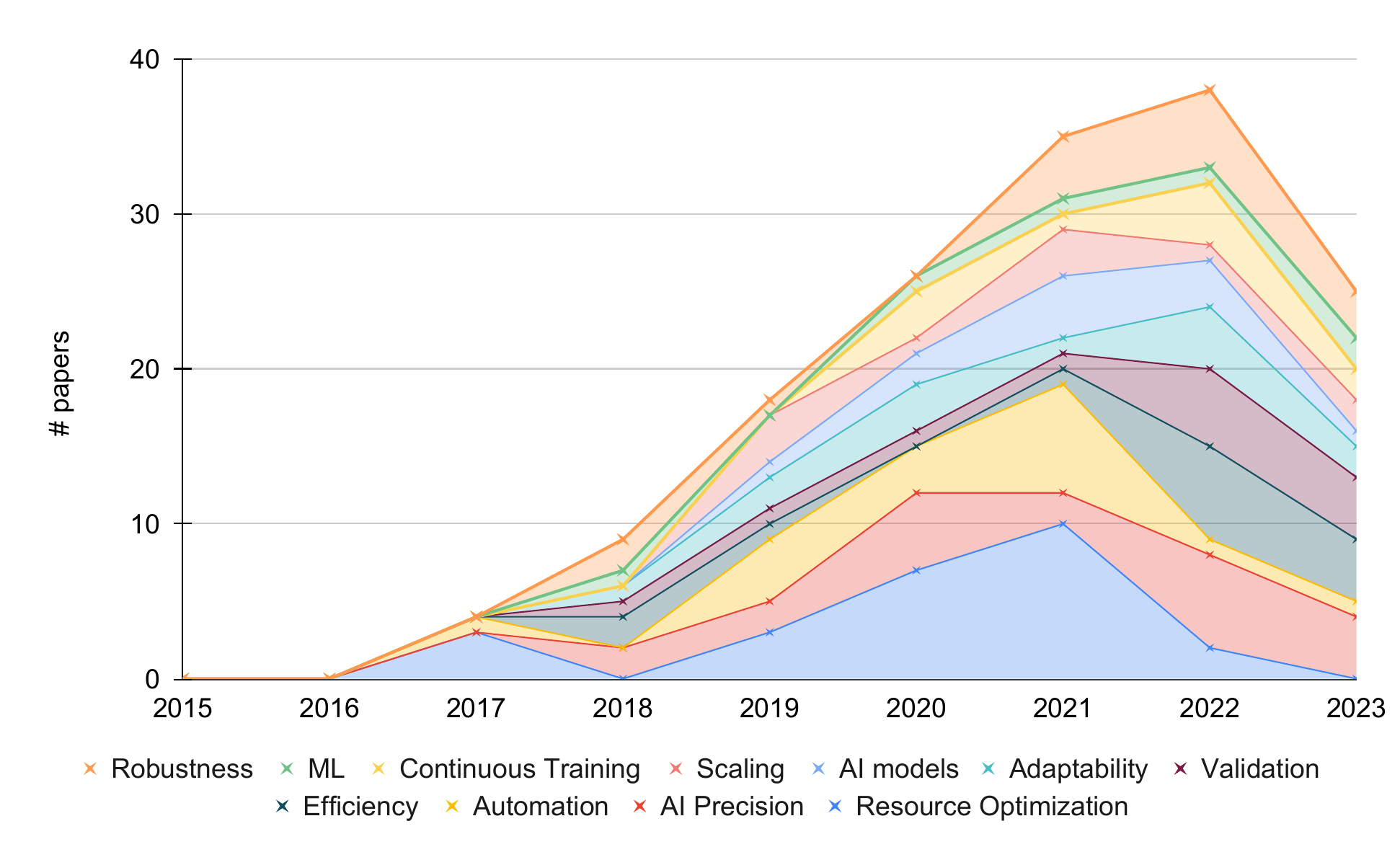}
    \caption{Future challenges per year}
    \vspace{-0.5cm}
    \label{fig:futhistyear}
\end{figure}



Among the 269 primary studies, 142 do not present a clear future challenge. When categorizing the challenges in the other studies, we identified different categories answering the questions \textit{What next?} or \textit{How?} We report them in \autoref{fig:fut}. 

 
Therein, \textit{Resource Optimization} and \textit{AI Precision} are the two aspects raising the most concern as future challenges in the AI4MS domain. The two aspects are mentioned in 25 and 21 papers, respectively. Combining such information shows that in most works, the authors are focused on presenting a minimum viable product (MVP) and leaving the optimization, mostly in the form of automation and resource optimization, for future challenges. The other important future challenges that have a relatively high number of mentions include \textit{Automation}, \textit{Efficiency}, \textit{Validation}, \textit{Adaptability}, and \textit{AI models}. Summarizing, future directions focus on improving the AI models and introducing automation mechanisms. 

Regarding the trend of the proposed future challenges (shown in Figure \ref{fig:futhistyear}), each of the top themes, e.g., \textit{Resource Optimization} and \textit{Automation}, have sharp increases in number of papers from 2019 to 2021. However, these topics decreased in numbers from 2021 to 2022, hence such challenges had received further investigation.  Vice versa, other challenges have caught the attention of the researchers. This includes \textit{AI precision}, \textit{Validation}, and \textit{Efficiency}, whose frequencies increased from 2021 to 2022.



\section{Multidimensional Analysis}
\label{sec:sa_qa_ai}
\begin{figure}
    \centering
    \includegraphics[width=0.8\linewidth]{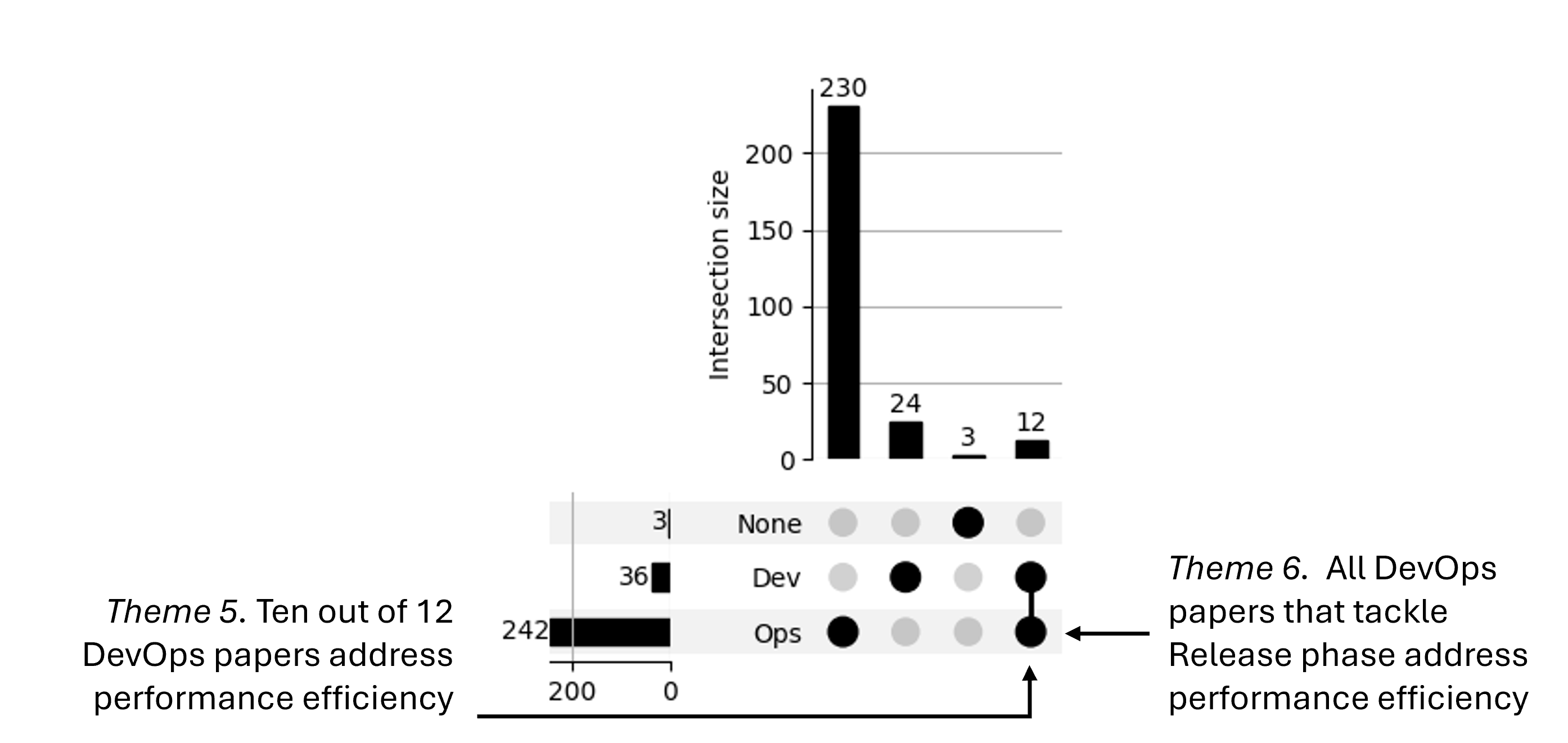}
    \caption{An Upset plot of the amount of papers addressing the Dev stage, the Ops stage or both stages}
    \label{fig:upsetPlotDevOps}
\end{figure}

\begin{figure}
    \centering
    \includegraphics[width=0.9\linewidth]{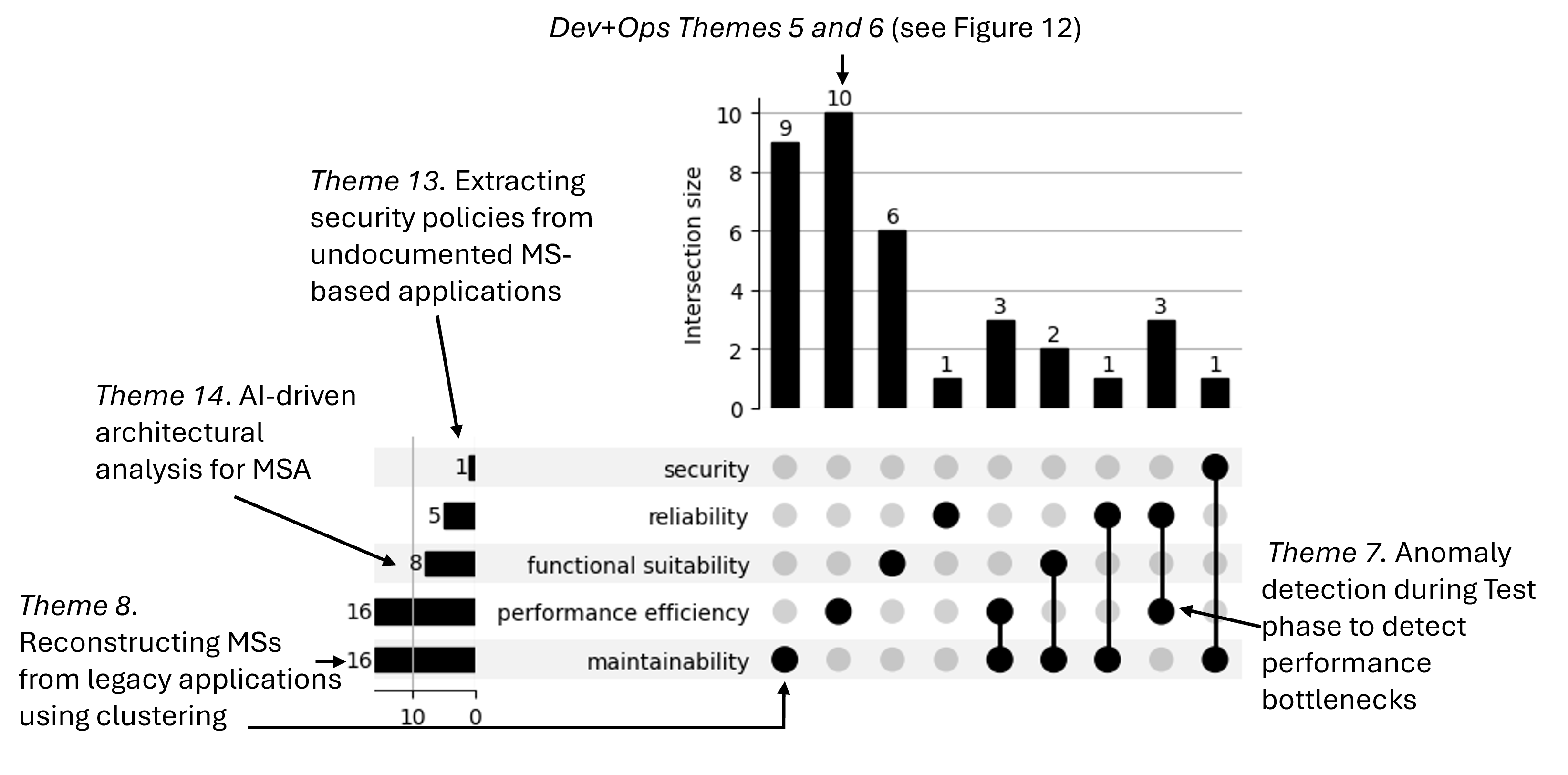}
    \caption{An Upset plot of the amount of papers addressing a specific intersection of QAs during the Dev phases (Plan, Code, Test, Release)}
    \label{fig:upsetplotDev}
\end{figure}

\begin{figure}
    \centering
    \includegraphics[width=1.1\linewidth]{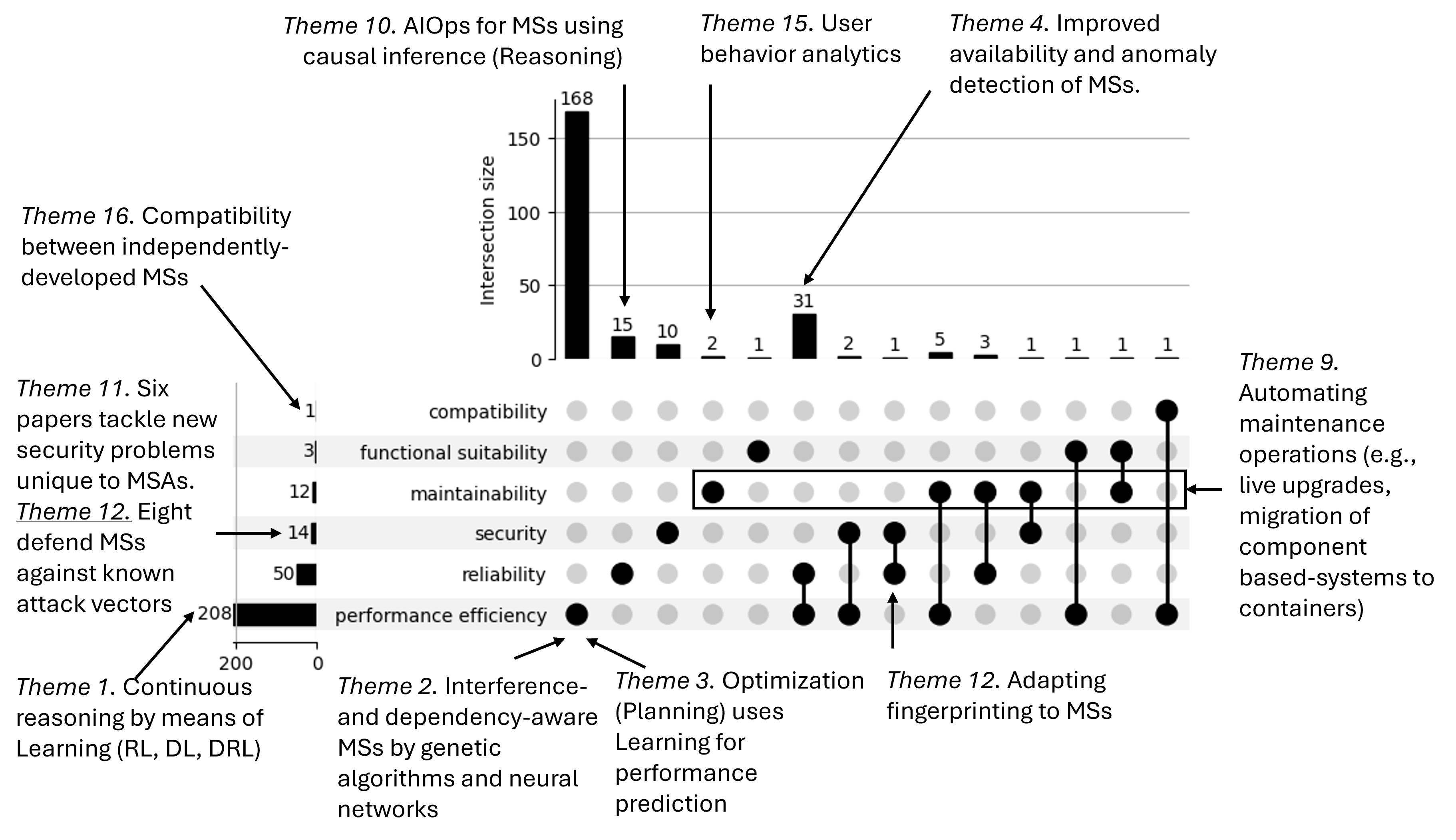}
    \caption{An Upset plot of the amount of papers addressing a specific intersection of QAs during the Ops phases (Deploy, Monitor, Operate)}
    \label{fig:upsetplotOps}
\end{figure}

While the data for each RQ can provide relevant insights individually, we also analyzed the combined data of several RQs for additional depth. 
Table~\ref{tab:SA_QA} shows the weight of research for each combination of DevOps phase and QA.
For each combination, we studied the extracted rationale and AI techniques to identify commonalities in research topics and approach. We found that distinct themes appear in the Dev stage (Plan, Code, Test, Release), Ops stage (Deploy, Operate, Monitor) and the full DevOps lifecycle. We identified 4 themes in the Dev stage (see Figure~\ref{fig:upsetplotDev}), 10 themes in the Ops stage (see Figure~\ref{fig:upsetplotOps}) and 2 performance efficiency themes in the full DevOps lifecycle (see Figure~\ref{fig:upsetPlotDevOps}) . 

\begin{table}[ht]
\centering
\begin{tabular}{lccccccc}
\toprule
& \textbf{Operate} & \textbf{Monitor} & \textbf{Deploy} & \textbf{Plan} & \textbf{Test} & \textbf{Code} & \textbf{Release} \\
\midrule
\textbf{Performance efficiency} & 147 & 69 & 69 & 4 & 4 & 2 & 6 \\
\textbf{Reliability} & 26 & 25 & 9 & 0 & 3 & 0 & 0 \\
\textbf{Maintainability} & 6  & 7 & 3 & 12 & 1 & 3 & 0 \\
\textbf{Security} & 6 & 10 & 2 & 1 & 0 & 0 & 0 \\
\textbf{Functional suitability} & 2 & 2 & 0 & 3 & 2 & 4 & 0 \\
\textbf{Compatibility} & 1 & 0 & 0 & 0 & 0 & 0 & 0 \\
\bottomrule
\end{tabular}
\caption{This table shows the amount of papers that cover a particular QA and DevOps phase. Papers that address multiple QAs or DevOps phases are counted multiple times.}
\vspace{-0.5cm}
\label{tab:SA_QA}
\end{table}

We present these 16 themes per QA and then per Ops (Deploy, Operate, Monitor) and Dev (Plan, Test, Code, Release) stage.


\textbf{Performance Efficiency:}
As shown by Table~\ref{tab:SA_QA} and  Figure~\ref{fig:upsetplotOps}, the bulk of papers improve \emph{performance efficiency during the Deploy, Operate and Monitor phases}. 
We identified several themes by studying the extracted rationale and AI techniques of these papers.

\textit{Theme 1: ML-based continuous reasoning.} Some AI techniques are very suitable for continuous reasoning on vast data dimensions and data sizes (e.g. analysis of QoS parameters, service and resource parameters), hereby also optimizing service selection, resource allocation and service placement strategies at run-time~\ref{SP143}. Moreover, \ref{SP172},~\ref{SP177} and~\ref{SP196} underscore the necessity for continuous reasoning. As stated by~\ref{SP143}, RL and DL are often used for this purpose, and  combined into Deep Reinforcement Learning (DRL) because of their respective ability of dynamic decision-taking and automated feature acquisition.

\textit{Theme 2: Interference- and dependency-aware scheduling of MSs using genetic algorithms and neural networks.} Many performance-only papers argue that during resource allocation and service placement, there is a need to account for (i) resource or availability interference between MSs \ref{SP17},~\ref{SP50},~\ref{SP243},~\ref{SP266}, very often using a genetic algorithm and (ii) inter-dependencies and associated call graphs~\ref{SP22},~\ref{SP130},~\ref{SP236},~\ref{SP209}, typically using a neural network. E.g., \ref{SP243} is a novel example of resource interference; it improves the performance efficiency of the Kubernetes scheduler with a genetic algorithm that places containers with shared library dependencies on the same node, hereby reducing resource usage due to the container library sharing mechanism. As an example of dependency management, \ref{SP130} feeds into the aforementioned continuous reasoning capabilities a neural network to predict the performance impacts and back log pressures that different MS inter-dependencies may cause in a production cloud system. 
    %
    
\textit{Theme 3: Optimization uses ML for performance prediction.} As stated in Section~\ref{sec:ai}, optimization from the Planning AI domain is massively used for improving performance efficiency of MSs. Optimization typically relies on the AI Learning techniques such as neural networks to predict the performance of particular configurations and/or resource allocation parameters in the search space. Specific to MSs, the integration of optimization techniques with these ML models addresses load balancing, resource utilization, autoscaling, risk management, and energy efficiency, ultimately leading to more robust and efficient cloud and microservice architectures.

\textit{Theme 4: Improved availability and anomaly detection of MSs.} As already elaborated in Section \ref{sec:qa}, 29 out of 34 papers that tackle both performance efficiency and reliability, improve  availability either by using multi-objective optimization methods or reducing  service downtime by means of anomaly detection, and do so exclusively during the Monitor or Operate phases.

With respect to \emph{performance efficiency during the early Dev phases} (Plan, Code, Test, Release), the following themes could be identified. 

\textit{Theme 5: Full DevOps lifecycle approaches mainly address performance efficiency}.  Out of 16 papers that address performance efficiency during the Dev stage, 10 papers address both Dev and Ops stages hereby representing 84\% of all the 12 papers that cover both stages (cf. Figure ~\ref{fig:upsetPlotDevOps}).

\textit{Theme 6: DevOps papers that tackle the Release phase all address performance efficiency}. No other QA than performance efficiency is addressed during the Release phase 
(cf. Table~\ref{tab:SA_QA}).  We found 6 papers that focus on the Release phase while also addressing the entire Ops stage (e.g., "Release, Deploy, Operate, Monitor"), i.e. representing 60\% of all the performance efficiency papers covering both stages. This consistent grouping of Release phase and the entire Ops stage reflects a streamlining of the CI/CD process where the release of code is closely followed by its deployment to production systems, supporting faster delivery cycles. Then there are 10 performance papers in the other Dev phases, 4 of which focus on performance efficiency only, 3 on the combination of~\emph{reliability and performance efficiency}, and 3 on \emph{maintainability and performance efficiency}.     

\textit{Theme 7: Anomaly detection during the Test phase.} As a common rationale, we could identify that 2 out of 3 \emph{reliability and performance efficiency} papers use anomaly detection during the Test phase to find performance bottlenecks in MS-based applications.

\textbf{Maintainability:}
\textit{Theme 8: Reconstructing MSs from legacy applications using clustering.} All the \emph{maintainability and performance efficiency} papers focus on optimizing the re-partitioning of monolithic applications into MSs to achieve the best performance. \ref{SP159} and~\ref{SP217} implement such re-partitioning during the Plan phase, whereas~\ref{SP136} provides a completely automated system for web applications that covers the Code, Deploy and Operate phases. It decomposes the application code into MSs, and deploys and auto-scales these with performance efficiency in mind.
Similar findings can be drawn for \emph{all maintainability papers that tackle the early Dev phases} (with or without a performance requirement). Out of 16 such papers, reconstructing MSs from legacy applications is the topic of 14 papers. However, similar work as the aforementioned~\ref{SP136} that covers Dev and Ops phase does not exist. There are only two other approaches~\ref{SP273},~\ref{SP122} that are executed during the Code phase, but they do not generate code artifacts. 
Finally, as already noted by Saucedo et al.~\cite{saucedo2024Clustering}, clustering and unsupervised ML in general is commonly used as the primary technique for supporting migration from monolithic applications to MSs. This is because these techniques allow inferring useful results from existing data sources without needing to label training data with explicit features based on prior or privileged knowledge~\ref{SP216}. Existing data sources include network metadata~\ref{SP216}, syntactic and semantic properties of object-oriented programs and databases~\ref{SP273},~\ref{SP246},~\ref{SP248}, and logs of non-functional metrics for determining appropriate units of resource allocation and service scaling~\ref{SP247},~\ref{SP136}. 

\textit{Theme 9. Automating maintenance operations during Ops stage.} With respect to \emph{maintainability during the later Ops phases}, 10 out of 12 papers \emph{combine the maintainability QA with another QA}, hereby employing a wide range of AI techniques. The common rationale that binds this work is automating complex tasks and reducing manual intervention, with a particular focus on better performance efficiency (5 out of 10) or reliability (3 out of 10) of MSs. 

\textbf{Reliability:}
\textit{Theme 10: AIOps for MSs using causal inference.} Out of 50 approaches that focus on \emph{reliability} during the Ops phases, there are only 15 papers that focus exclusively on reliability, but for these papers, improving AIOps for MSs is a common trend in the extracted rationale, especially for papers published in 2023. AIOps is an approach to collect, analyze, and detect patterns in cloud and infrastructure data, thus predicting future usage, failures, and improving the management and resilience of complex IT environments~\ref{SP144},\hl{~\ref{SP127}},~\ref{SP230}. Unique to MSs in AIOps is the use of neural networks, DL and causal inference to handle the complexity and dynamism of MSAs, which involve numerous interdependent services with complex spatial states and hundreds of metrics. Specific topics that are frequently handled include (1) log-based anomaly detection and fault localization~\ref{SP154},~\ref{SP160},~\ref{SP242}, (2) selecting appropriate metrics as features in supervised ML~\ref{SP200}, (3) causal dependency learning to observe error propagation~\ref{SP144},~\ref{SP199}, and (4) proactive and self-learning systems that become stronger upon faults rather than deteriorating~\ref{SP140},~\ref{SP198},~\ref{SP98}. Interestingly, causal inference from the Reasoning domain appears to be better than ML in this space~\ref{SP66}. As an example, both~\ref{SP199} and~\ref{SP144} use a specific technique called interventional causal learning.

%
%
%
%

\textbf{Security:}
We identified two research themes for the papers that focus on security in the Ops stage. Note that for 12 out of 14 papers, the security improvements are implemented during \emph{the Monitor or Operate phases}.

\textit{Theme 11: Tackling new security problems unique to MSAs.} One trend is the detection and mitigation of new security anomalies and attacks that are due to unique properties of MSAs. There are 2 papers that use neural networks to profile the abnormal application-level behavior of MSs from highly distributed, heterogeneous and unstructured data~\ref{SP49},~\ref{SP168}. Other papers use AI to implement self-adaptive anomaly detection solutions that can cope with the dynamic and evolving nature of microservice environments~\ref{SP21},~\ref{SP161} or that can be automatically applied to different applications~\ref{SP163}.  Finally, \ref{SP151} presents a DRL-based scheduler to reduce lateral movement of attackers across the network of a Kubernetes cluster. The scheduler aims to determine subsets of applications that have similar microservice call chains and then exclusively place containers of the identified applications together on set of nodes in line with the identified chain patterns.

\textit{Theme 12: Defending MSs against known attack vectors.}
The bulk of the security papers focus on defending MSs against known attack vectors or adapting existing defenses to MSs. Attack vectors include data breaches and denial-of-service attacks~\ref{SP161}, malicious threat patterns and zero-day vulnerabilities in containers~\ref{SP11}, IoT network attacks~\ref{SP137}, password guess attacks~\ref{SP234}. Adapted security defenses include fingerprinting~\ref{SP90}. 

\textit{Theme 13. Extracting security policies from undocumented MS-based applications.} We only identified 1 paper that targets \emph{security during the Dev stage}~\ref{SP216}. This work provides a three-fold mechanism operating during the \textit{Plan} phase: first,  a reconstruction of inter-MS interactions from undocumented MS-based applications is performed; then the extracted interactions are classified as normal or abnormal; and finally appropriate access control policies are defined accordingly.

\textbf{Less Prevalent QAs:}
Although there are not so many \emph{functional suitability} papers, we could distill a common research theme.

\textit{Theme 14: AI-driven architectural analysis for MSAs}. For the \emph{Dev} stage, all 8 functional suitability papers support AI-driven architectural analysis of MSAs. These papers contribute with approaches that help software designers identify suitable boundaries and granularity for MSs using various AI techniques such as Natural Language Processing~\ref{SP57}~\ref{SP122}, clustering~\ref{SP250}, generative AI~\ref{SP235}, DL~\ref{SP84}, and recommender systems~\ref{SP76}.  
Another work supports test case prioritization by quantifying the invocation weight of MSs using a recommender system based on Page Rank~\ref{SP186}. Finally, there is an agile approach that operates at the intersection of functional suitability and maintainability to quickly grasp the impact of new requirements on a code base using classification and neural networks~\ref{SP8}. 

\textit{Theme 15: User behavior analytics.} There are only two pure functional suitability papers that tackle the \emph{Ops} phase. Both papers support user behavior analytics by monitoring user interactions with MSs~\ref{SP171}~\ref{SP149}. 

\textit{Theme 16: Compatibility between independently developed MSs during Operate.} The single paper addressing \emph{compatibility} during the \emph{Operate} phase improves the composition and co-existence of independently developed MSs~\ref{SP269}.

\section{Discussion}
\label{Discussion}
Figure~\ref{fig:publications-year} clearly shows the growing application of AI in the field of MSs. Significantly, most publications in AI4MS —despite having its roots in the industry— involve academics. Our decision to only consider peer-reviewed publications may be pertinent in this regard given that businesses typically present their outcomes through speeches and blog posts rather than peer-reviewed papers.

The most used AI techniques are ML and its different incarnations, but also more specific techniques such as optimization and anomaly detection are heavily used.  

The primary use of AI until now has been to increase performance efficiency not just in the Operate phase but also during monitoring and deployment. 
Reliability is typically taken into account during the same phases as performance efficiency and has led to improved AIOps for MSs. Instead, security-related methods concentrate on the monitoring stage. Significantly, almost all of the techniques emphasize the Ops phases, with almost none focusing on the Dev phases (no approach at all considers Build). However, a notable exception to this concerns maintainability papers that implement automated migration of legacy applications to MSs during the Dev stages. Reference \ref{SP136} is however the only automated tool for automatic refactoring of a monolithic artifact into code for MSs and Ops artifacts for service placement and resource allocation, but this approach is limited to web applications.  


Our results highlight a few gaps in the literature. To put the list below into context, consider that the first peer-reviewed works on AI4MS were published only in 2017; therefore, research concentrated on the simplest problems. Also, the newest trends may currently be considered only in industry, hence not yet disseminated via refereed publications. The main identified gaps are described hereafter.

\smallskip \noindent \textbf{AI in Dev phases}. Most approaches focus on Deploy, Operate and Monitor (cf.~Fig.~\ref{fig:devops}). We expect AI to be able to play a major role also in the Plan (e.g., automatic requirement analysis), Code (e.g., automatic refactoring tools that also generate deployment artifacts) and Test (e.g., automatic test case generation) phases.

\smallskip \noindent \textbf{Portability, compatibility and usability}. As shown in Section~\ref{sec:qa}, no research in our sample aims to improve these QAs, but for a single paper targeting compatibility. Several open research questions can be derived from this observation. First, while MSs are inherently more portable and interoperable (the 1st sub-charac\-teris\-tic of compatibility~\cite{InternationalOrganizationForStandardization2011}), it is not known whether AI techniques such as natural language processing, expert systems and generative AI (see below for a more detailed discussion of generative AI) can improve the level of automation in vendor-agnostic model-driven configuration methods such as TOSCA~\cite{Soldani2021_muTosca}. Second, coexistence (the 2nd sub-charac\-teris\-tic of compatibility~\cite{InternationalOrganizationForStandardization2011}), which can be interpreted as the desire to reduce dysfunctional emergent behavior, caused by feature interactions between MSs, can definitively be improved by means of several AI techniques such as (a) genetic and evolutionary algorithms, (b) multi-agent systems, reinforcement learning, or a combination of both, (c) anomaly detection. Third, while AI-assisted selection of MSs for improving functional suitability has been marginally studied (e.g., \ref{SP57}), there is a lack of understanding of to which extent AI-assisted  user interface design must be done differently in the era of MSs. 

\smallskip \noindent \textbf{AI for security}. While security is nowadays a main concern, it is clear from \autoref{tab:SA_QA} that only a few works consider it, and they are concentrated in the monitoring and operate phases, to detect anomalies or ongoing attacks. We believe AI can contribute much more to tackling security issues in MSs, and such contributions can take place in most phases. Also, as discussed in Section~\ref{sec:qa} approaches for security do not improve other QAs, hence security calls for dedicated approaches and techniques.

\smallskip \noindent \textbf{Generative AI for DevOps}. A relevant instance of the use of AI in Dev phases, and in particular in the Code phase, concerns exploiting generative AI such as ChatGPT to assist programmers in code writing. Such approaches have born recently since ChatGPT was released towards the end of 2022,  and are starting to be applied to MSs as well\footnote{\url{https://frends.com/video/creating-a-microservice-with-chatgpt-in-2-minutes}}. We believe such a research direction will gain interest in the future, hence we expect AI keywords such as Chatbot, occurring only in one of the primary studies we consider~\ref{SP235}, will gain emphasis. Similar approaches could also be used to generate other artifacts, such as specifications, tests or documentation, and more in general to provide a natural language interface to tools.   
As for Ops, continuous monitoring, anomaly detection, and self-healing might be dominated by self-learning and generative AI tools in the near future. 

\smallskip \noindent \textbf{Explainability}. Most of the surveyed AI techniques are not \enquote{explainable by design}, meaning that, despite they support MSs in their DevOps life-cycle, they are not providing explanations of why this is the case.
AI techniques can indeed be used to, e.g., determine performance/functional anomalies, identify the root causes of failures, or detect security leaks/intrusions.
At the same time, associating identified issues to why they are considered so would help DevOps engineers in troubleshooting them and patching MSs to avoid such issues to happen again in the future, also focusing only on true positives. 
This hence calls for AI techniques that support MSs in their DevOps life-cycle while also being \enquote{explainable by design}, much in the same way as the need for explainability is nowadays recognized in AI \cite{Guidotti2019_Explainability}. Among primary studies, ~\ref{SP81} and~\ref{SP130} studies correspond to explainable AI.


\section{Threats to Validity}
\label{Threats}
The results of an SMS may be subject to validity threats, mainly concerning the correctness and completeness of the survey. We follow the guidelines for identifying the threats to validity in secondary studies in the software engineering domain proposed by Ampatzoglou et al. \cite{ampatzoglou2019identifying}. We discuss them below.



\textbf{Study Selection Validity.} In this study, we strictly follow the established and commonly accepted SMS guidelines in terms of the search strategy, review protocol, and the data extraction process~\cite{Kitchenham2007}. By doing so, we significantly reduced the threats to the initial search and study filtering processes in the secondary study planning phase. To do so, the search string was formulated to include keywords identified from research questions and diversified using synonyms. However, though most of the publications are covered by the initial search, potential limitations on the search string may still evoke issues, which results in missing key studies. To mitigate the search limitations and extend the coverage of studies, we conducted snowballing, where we reviewed all the references listed in the selected studies and all the papers that reference the selected ones. 
Snowballing was recursively applied to papers coming from snowballing as well. As it was likely that the snowballing activity could continue for an excessively long period, the snowballing activity ceased at the end of January 2024.
The inclusion and exclusion criteria were defined to assist the study selection. The criteria aligned with the paper's goal and research questions and the guidelines recommended by Petersen et al.~\cite{PETERSEN20151}. The selection process prescribed that at least two authors conducted the study selection independently, with a third author involved in the discussion to resolve any disagreement. 


\textbf{Data Validity.} The data extraction process is a similar procedure where two authors conducted an iterative analytic process driven by the open coding method to identify the classification schema. 
For certain categories, we adopted publicly available standards. For example, to answer RQ2, we adopt the ISO/IEC 25010 software quality model, which is commonly acknowledged as the cornerstone of a product quality evaluation system and determines the quality characteristics considered when evaluating software quality. By adopting such open standards, we shall avoid potential disagreement and bias, as well as guarantee the correctness of the collected data. For the data analysis process, thanks to the pre-defined categories, the extracted results can be easily summarized and displayed in the form of bar charts. On the other hand, publication bias is also a potential threat to data validity, where methods, techniques, and usage goals from companies are not included sufficiently due to the focus on peer-reviewed papers as well as confidential policies. Such a perspective can be further investigated by analyzing grey literature and industrial surveys in future studies. 

\section{Conclusion}
\label{Conclusion}

In this paper, we conducted a systematic mapping study on the use of AI in the life-cycle of MS systems. Based on the selected 269 primary studies, we focus on understanding, in the area of MSs, which AI technologies are used, in which domain and according to which rationale, namely which software quality attributes the AI technologies aim to improve, and in which DevOps phases. 

The results show that AI4MS is a trendy area, with increasing numbers of studies 
in many application areas. 
The main outcomes are: 1) while the main application area is, of course, IT, manufacturing is also starting to attract interest; 2) the main rationale is improving performance efficiency and reliability in Ops phases, while surprisingly Dev phases are rarely considered, and QAs such as Portability and Usability are not considered at all; 3) current research focuses on building the minimum viable product showcasing some approach, with optimization and automation left for future work; 4) a multi-dimensional analysis identifies 16 research themes that include among others the use of deep reinforcement learning for performance efficiency, AIOps for MSs, tackling new security problems unique to MSAs, and adapting existing security techniques to MSs.

This paper provides insights on AI4MS, by keeping the discussion at a high level mainly due to the quite considerable amount of currently available/selected studies. Future work will include delving more into the details of sub-areas of AI4MS, which can be achieved by selecting subsets of the already selected studies based on additional selection criteria. In particular, we plan to analyse how AIOps and MLOps are currently used in the life-cycle of MSs. Other than going into the details, by narrowing the focus to AIOps and MLOps, it would become manageable to complement our analysis of peer-reviewed literature with grey literature, to shed light on both the state-of-the-art and state-of-practice on the topic. This is part of our future work.

On another front, we plan to complement the results presented in this study by analysing the dual situation, namely how MSs are used to support the design, development, and operation of AI systems.

\section*{Acknowledgements}

Ivan Lanese has been partially supported by French ANR project SmartCloud ANR-23-CE25-0012.  Shahrzad Pour has been partially supported by the European Commission fundings through DeployAI (Grant No:101146490) and BIPED (Grant No: 101139060) projects.
Jacopo Soldani has been partly supported by the project FREEDA (CUP: I53D23003550006), funded by PRIN (MUR, Italy) and Next Generation EU. 
This research is partially funded by the Research Fund KU Leuven.








\bibliography{sn-bibliography}
\begin{appendices}
\newpage
\section{Primary Studies}\label{app:ps}
{\small
\begin{enumerate}[labelindent=-5pt,label={[SP}{\arabic*]}]

\smallskip\item \label{SP1} 
S. Taherizadeh, V. Stankovski, M. Grobelnik, A capillary computing architecture for dynamic internet of things: Orchestration of microservices from edge devices to fog and cloud providers, Sensors 18 (2018).
\smallskip\item \label{SP2}
J. Lv, M. Wei, Y. Yu, A container scheduling strategy based on machine learning in microservice architecture, in: 2019 IEEE International Conference on Services Computing (SCC).
\smallskip\item \label{SP3}
Y. Li, L. Chen, D. Zeng, L. Gu, A customized reinforcement learning based binary offloading in edge cloud, in: 2020 IEEE 26th International Conference on Parallel and Distributed Sys- tems (ICPADS).
\smallskip\item \label{SP5}
B. Magableh, M. Almiani, A deep recurrent q network towards self-adapting distributed microservice architecture, Software: Practice and Experience 50 (2020).
\smallskip\item \label{SP6}
A. Goli., N. Mahmoudi., H. Khazaei., O. Ardakanian., A holistic machine learning-based autoscaling approach for microservice applications, in: Proceedings of the 11th International Conference on Cloud Computing and Services Science - CLOSER,, INSTICC, SciTePress, 2021.
\smallskip\item \label{SP7}
J. Cui, P. Chen, G. Yu, A learning-based dynamic load balancing approach for microservice systems in multi-cloud environment, in: 2020 IEEE 26th International Conference on Parallel and Distributed Systems (ICPA-DS).
\smallskip\item \label{SP8}
D. Russo, V. Lomonaco, P. Ciancarini, A machine learning approach for continuous development, in: P. Ciancarini, S. Litvinov, A. Messina, A. Sillitti, G. Succi (Eds.), Proceedings of 5th International Conference in Software Engineering for Defence Applications, Springer International Publishing, Cham, 2018.
\smallskip\item \label{SP9}
M. Caporuscio, M. De Toma, H. Muccini, K. Vaidhyana-than, A machine learning approach to service discovery for microservice architectures, in: S. Biffl, E. Navarro, W. L\"owe, M. Sirjani, R. Mirandola, D. Weyns (Eds.), Software Architecture, Springer International Publishing, Cham, 2021.
\smallskip\item \label{SP10}
\begin{sloppypar}
G. S. Siriwardhana, N. De Silva, L. S. Jayasinghe, L. Vithanage, D. Kasthurirathna, A network science-based approach for an optimal microservice governance, in: 2020 2nd International Conference on Advancements in Computing (ICAC), volume 1.
\end{sloppypar}
\smallskip\item \label{SP11}
S. Kamthania, A novel deep learning rbm based algorithm for securing containers, in: 2019 IEEE International WIE Conference on Electrical and Computer Engineering (WIECON-ECE).
\smallskip\item \label{SP12}
M. G. Khan, J. Taheri, M. A. Khoshkholghi, A. Kassler, C. Cartwright, M. Darula, S. Deng, A performance modelling approach for sla-aware resource recommendation in cloud native network functions, in: 2020 6th IEEE Conference on Network Softwarization (NetSoft).
\smallskip\item \label{SP13}
C. Liu, Z. Cai, B. Wang, Z. Tang, J. Liu, A protocol-independent container network observability analysis system based on ebpf, in: 26th IEEE International Conference on Parallel and Distributed Systems, ICPADS 2020, Hong Kong, December 2-4, 2020, IEEE, 2020.
\smallskip\item \label{SP14}
S. Agarwal, M. A. Rodriguez, R. Buyya, A reinforcement learning approach to reduce serverless function cold start frequency, in: 2021 IEEE/ACM 21st International Symposium on Cluster, Cloud and Internet Computing (CCGrid).
\smallskip\item \label{SP15}
Z. He, P. Chen, X. Li, Y. Wang, G. Yu, C. Chen, X. Li, Z. Zheng, A spatiotemporal deep learning approach for unsupervised anomaly detection in cloud systems, IEEE Transactions on Neural Networks and Learning Systems (2020).
\smallskip\item \label{SP16}
L. Toka, G. Dobreff, B. Fodor, B. Sonkoly, Adaptive {AI}-based auto-scaling for {K}ubernetes, in: 2020 20th IEEE/ ACM International Symposium on Cluster, Cloud and Internet Computing (CCGRID).
\smallskip\item \label{SP17}
N. Cruz Coulson, S. Sotiriadis, N. Bessis, Adaptive microservice scaling for elastic applications, IEEE Internet of Things Journal 7 (2020).
\smallskip\item \label{SP18}
K. Fu, W. Zhang, Q. Chen, D. Zeng, M. Guo, Adaptive resource efficient microservice deployment in cloud-edge continuum, IEEE Transactions on Parallel and Distributed Systems 33 (2022).
\smallskip\item \label{SP19}
R. A. Addad, D. L. C. Dutra, T. Taleb, H. Flinck, Ai-based network-aware service function chain migration in 5g and beyond networks, IEEE Transactions on Network and Service Management 19 (2022).
\smallskip\item \label{SP20}
H. Sami, H. Otrok, J. Bentahar, A. Mourad, AI-Based Resource Provisioning of IoE Services in 6G: A Deep Reinforcement Learning Approach, IEEE Transactions on Network and Service Management 18 (2021).
\smallskip\item \label{SP21}
M. -O. Pahl, F. -X. Aubet, All eyes on you: Distributed multi-dimensional iot microservice anomaly detection, in: 2018 14th International Conference on Network and Service Management (CNSM), 2018.
\smallskip\item \label{SP22}
X. Hou, C. Li, J. Liu, L. Zhang, S. Ren, J. Leng, Q. Chen, M. Guo, Alphar: Learning-powered resource management for irregular, dynamic microservice graph, in: 2021 IEEE International Parallel and Distributed Processing Symposium (IPDPS).
\smallskip\item \label{SP23}
M. Jin, A. Lv, Y. Zhu, Z. Wen, Y. Zhong, Z. Zhao, J. Wu, H. Li, H. He, F. Chen, An anomaly detection algorithm for microservice architecture based on robust principal component analysis, IEEE Access 8 (2020).
\smallskip\item \label{SP24}
Y. Zuo, Y. Wu, G. Min, C. Huang, K. Pei, An intelligent anomaly detection scheme for micro-services architectures with temporal and spatial data analysis, IEEE Transactions on Cognitive Communications and Networking 6 (2020).
\smallskip\item \label{SP25}
S. Nedelkoski, J. Cardoso, O. Kao, Anomaly detection and classification using distributed tracing and deep learning, in: 2019 19th IEEE/ACM International Symposium on Cluster, Cloud and Grid Computing (CCGRID).
\smallskip\item \label{SP26}
Q. Du, T. Xie, Y. He, Anomaly detection and diagnosis for container-based microservices with performance monitoring, in: J. Vaidya, J. Li (Eds.), Algorithms and Architectures for Parallel Processing, Springer International Publishing, Cham, 2018.
\smallskip\item \label{SP27}
M. Lin, J. Xi, W. Bai, J. Wu, Ant colony algorithm for multi-objective optimization of container-based microservice scheduling in cloud, IEEE Access 7 (2019).
\smallskip\item \label{SP28}
G. Baye, F. Hussain, A. Oracevic, R. Hussain, S. Ahsan Kazmi, Api security in large enterprises: Leveraging machine learning for anomaly detection, in: 2021 International Symposium on Networks, Computers and Communications (ISNCC).
\smallskip\item \label{SP29}
A. U. Gias, G. Casale, M. Woodside, Atom: Model-driven autoscaling for microservices, in: 2019 IEEE 39th International Conference on Distributed Computing Systems (ICDCS).
\smallskip\item \label{SP30}
I. Prachitmutita, W. Aittinonmongkol, N. Pojjanasuksakul, M. Supattatham, P. Padungweang, Auto-scaling microservices on iaas under sla with cost-effective framework, in: 2018 Tenth International Conference on Advanced Computational Intelligence (ICACI).
\smallskip\item \label{SP31}
\begin{sloppypar}
N. Chondamrongkul, J. Sun, I. Warren, Automated planning for software architectural migration, in: 2020 25th International Conference on Engineering of Complex Computer Systems (ICECCS), 2020.
\end{sloppypar}
\smallskip\item \label{SP32}
A. Samanta, Y. Li, F. Esposito, Battle of microservices: Towards latency-optimal heuristic scheduling for edge computing, in: 2019 IEEE Conference on Network Softwarization (NetSoft).
\smallskip\item \label{SP33}
S. B. Nath, S. Chattopadhyay, R. Karmakar, S. K. Addya, S. Chakraborty, S. K. Ghosh, Containerized deployment of micro-services in fog devices: a reinforcement learning-based approach, The Journal of Supercomputing 78 (2021).
\smallskip\item \label{SP34}
\begin{sloppypar}
R. C. Chiang, Contention-aware container placement strategy for docker swarm with machine learning based clustering algorithms, Cluster Computing (2020).
\end{sloppypar}
\smallskip\item \label{SP35}
M. De Sanctis, H. Muccini, and K. Vaidhyanathan. Data-driven adaptation in microservice-based iot architectures, in 2020 IEEE International Conference on Software Architecture Companion (ICSA-C). IEEE, 2020.
\smallskip\item \label{SP36}
R. Yu, S.-Y. Lo, F. Zhou, G. Xue, Data-driven edge resource provisioning for inter-dependent microservices with dynamic load, in: 2021 IEEE Global Communications Conference (GLOBECOM).
\smallskip\item \label{SP37}
N.-M. Dang-Quang, M. Yoo, Deep learning-based autoscaling using bidirectional long short-term memory for kubernetes, Applied Sciences 11 (2021).
\smallskip\item \label{SP38}
R. Li, M. Du, H. Chang, S. Mukherjee, E. Eide, Deepstitch: Deep learning for cross-layer stitching in microservices, in: Proceedings of the 2020 6th International Workshop on Container Technologies and Container Clouds, WOC’20, Association for Computing Machinery, New York, NY, USA, 2021.
\smallskip\item \label{SP39}
S. Wang, Y. Guo, N. Zhang, P. Yang, A. Zhou, X. Shen, Delay-aware microservice coordination in mobile edge computing: A reinforcement learning approach, IEEE Transactions on Mobile Computing 20 (2021).
\smallskip\item \label{SP40}
S. Y. Shah, Z. Yuan, S. Lu, P. Zerfos, Dependency analysis of cloud applications for performance monitoring using recurrent neural networks, in: 2017 IEEE International Conference on Big Data (Big Data).
\smallskip\item \label{SP41}
S. Wu, C. Denninnart, X. Li, Y. Wang, M. A. Salehi, Descriptive and predictive analysis of aggregating functions in server- less clouds: the case of video streaming, in: 2020 IEEE 22nd International Conference on High Performance Computing and Communications, IEEE 18th International Conference on Smart City, IEEE 6th International Conference on Data Science and Systems (HP- CC/SmartCity/DSS).
\smallskip\item \label{SP42}
V. Cortellessa, L. Traini, Detecting latency degradation patterns in service-based systems, in: Proceedings of the ACM/SPEC International Conference on Performance Engineering, ICPE ’20, Association for Computing Machinery, New York, NY, USA, 2020.
\smallskip\item \label{SP43}
J. Grohmann, S. Eismann, S. Elflein, J. V. Kistowski, S. Kounev, M. Mazkatli, Detecting parametric dependencies for performance models using feature selection techniques, in: 2019 IEEE 27th International Symposium on Modeling, Analysis, and Simulation of Computer and Telecommunication Systems (MASCOTS).
\smallskip\item \label{SP44}
J. Chou, E. Al-Masri, S. Kanzhelev, H. Fattah, Detecting security and privacy risks in microservices end-to-end communication using neural networks, in: 2021 IEEE 4th International Conference on Knowledge Innovation and Invention (ICKII).
\smallskip\item \label{SP45}
A. A. Khaleq, I. Ra, Development of qos-aware agents with reinforcement learning for autoscaling of microservices on the cloud, in: 2021 IEEE International Conference on Autonomic Computing and Self-Organizing Systems Companion (ACSOSC).
\smallskip\item \label{SP46}
C. Hou, T. Jia, Y. Wu, Y. Li, J. Han, Diagnosing performance issues in microservices with heterogeneous data source, in: 2021 IEEE Intl. Conf. on Parallel Distributed Processing with Applications, Big Data Cloud Computing, Sustainable Computing Communications, Social Computing Networking (ISPA/BDCloud/So-cialCom/Sustain- Com).
\smallskip\item \label{SP47}
H. Tian, X. Xu, T. Lin, Y. Cheng, C. Qian, L. Ren, M. Bilal, Dima: Distributed cooperative microservice caching for internet of things in edge computing by deep reinforcement learning, World Wide Web 25 (2022).
\smallskip\item \label{SP48}
M. Abdullah, W. Iqbal, F. Bukhari, A. Erradi, Diminishing returns and deep learning for adaptive cpu resource allocation of containers, IEEE Transactions on Network and Service Management 17 (2020).
\smallskip\item \label{SP49}
M. Ghorbani, F. F. Moghaddam, M. Zhang, M. Pourzandi, K. K. Nguyen, M. Cheriet, Distappgaurd: Distributed application behaviour profiling in cloud-based environment, in: Annual Computer Security Applications Conference, ACSAC ’21, Association for Computing Machinery, New York, NY, USA, 2021.
\smallskip\item \label{SP50}
H. Zhao, S. Deng, Z. Liu, J. Yin, S. Dustdar, Distributed redundant placement for microservice-based applications at the edge, IEEE Transactions on Services Computing 15 (2022).
\smallskip\item \label{SP51}
\begin{sloppypar}
A. Samir, C. Pahl, Dla: Detecting and localizing anomalies in containerized microservice architectures using markov models, in: 2019 7th International Conference on Future Internet of Things and Cloud (FiCloud).
\end{sloppypar}
\smallskip\item \label{SP52}
A. Samanta, J. Tang, Dyme: Dynamic microservice scheduling in edge computing enabled iot, IEEE Internet of Things Journal 7 (2020).
\smallskip\item \label{SP53}
K. Zeng, I. Paik, Dynamic service recommendation using lightweight {BERT}-based service embedding in edge computing, in: 2021 IEEE 14th International Symposium on Embedded Multicore/Many-core Systems-on-Chip (MCSoC).
\smallskip\item \label{SP54}
G. Orsini, W. Posdorfer, W. Lamersdorf, Efficient mobile clouds: Forecasting the future connectivity of mobile and iot devices to save energy and bandwidth, Procedia Computer Science 155 (2019) 121–128. The 16th International Conference on Mobile Systems and Pervasive Computing (MobiSPC 2019),The 14th International Conference on Future Networks and Communications (FNC-2019),The 9th International Conference on Sustainable Energy Information Technology.
\smallskip\item \label{SP55}
S. Wang, Z. Ding, C. Jiang, Elastic scheduling for microservice applications in clouds, IEEE Transactions on Parallel and Distributed Systems 32 (2021).
\smallskip\item \label{SP56}
H. Mohamed, O. El-Gayar, End-to-end latency prediction of microservices workflow on kubernetes: A comparative evaluation of machine learning models and resource metrics, volume 2020-January.
\smallskip\item \label{SP57}
M. França, C. Werner, Evaluating cloud microservices with director, in: 2019 IEEE International Conference on Service-Oriented System Engineering (SOSE).
\smallskip\item \label{SP58}
V. M. Mostofi, D. Krishnamurthy, M. Arlitt, Fast and efficient performance tuning of microservices, in: 2021 IEEE 14th International Conference on Cloud Computing (CLOUD).
\smallskip\item \label{SP59}
H. Qiu, S. S. Banerjee, S. Jha, Z. T. Kalbarczyk, R. K. Iyer, Firm: An intelligent fine-grained resource management framework for slo-oriented microservices.
\smallskip\item \label{SP60}
C. T. Joseph, J. P. Martin, K. Chandrasekaran, A. Kandasamy, Fuzzy reinforcement learning based microservice allocation in cloud computing environments, in: TENCON 2019 - 2019 IEEE Region 10 Conference (TENCON).
\smallskip\item \label{SP61}
D. C. Li, C.-T. Huang, C.-W. Tseng, L.-D. Chou, Fuzzy-based microservice resource management platform for edge computing in the internet of things, Sensors 21 (2021).
\smallskip\item \label{SP62}
C. Guerrero, I. Lera, C. Juiz, Genetic algorithm for multi-objective optimization of container allocation in cloud architecture, Journal of Grid Computing 16 (2018).
\smallskip\item \label{SP63}
\begin{sloppypar}
J. Kim, S. Ullah, D.-H. Kim, Gpu-based embedded edge server configuration and offloading for a neural network service, The Journal of Supercomputing 77 (2021).
\end{sloppypar}
\smallskip\item \label{SP64}
J. Park, B. Choi, C. Lee, D. Han, Graf: A graph neural network based proactive resource allocation framework for slo-oriented microservices, in: Proceedings of the 17th International Conference on Emerging Networking EXperiments and Technologies, CoNEXT ’21, Association for Computing Machinery, New York, NY, USA, 2021.
\smallskip\item \label{SP65}
R. S. Kannan, L. Subramanian, A. Raju, J. Ahn, J. Mars, L. Tang, Grandslam: Guaranteeing slas for jobs in microservices execution frameworks, in: Proceedings of the Fourteenth EuroSys Conference 2019, EuroSys ’19, Association for Computing Machinery, New York, NY, USA, 2019.
\smallskip\item \label{SP66}
\'Alvaro Brand\'on, M. Sol\'e, A. Hu\'elamo, D. Solans, M. S. P\'erez, V. Munt\'es-Mulero, Graph-based root cause analysis for service-oriented and microservice architectures, Journal of Systems and Software 159 (2020).
\smallskip\item \label{SP67}
M. Yan, X. Liang, Z. Lu, J. Wu, W. Zhang, Hansel: Adaptive horizontal scaling of microservices using bi-lstm, Applied Soft Computing 105 (2021).
\smallskip\item \label{SP68}
F. Rossi, V. Cardellini, F. L. Presti, Hierarchical scaling of microservices in kubernetes, in: 2020 IEEE International Con- ference on Autonomic Computing and Self-Organizing Systems (ACSOS).
\smallskip\item \label{SP69}
J. Chen, H. Huang, H. Chen, Informer: Irregular traffic detection for containerized microservices rpc in the real world, High-Confidence Computing 2 (2022).
\smallskip\item \label{SP70}
M. Zhu, H. Qu, J. Zhao, Instance expansion algorithm for micro-service with prediction, Electronics Letters 54 (2018).
\smallskip\item \label{SP71}
A. A. Khaleq, I. Ra, Intelligent autoscaling of microservices in the cloud for real-time applications, IEEE Access 9 (2021).
\smallskip\item \label{SP72}
D. Uzunidis, P. Karkazis, C. Roussou, C. Patrikakis, H. C. Leligou, Intelligent performance prediction: The use case of a hadoop cluster, Electronics 10 (2021).
\smallskip\item \label{SP73}
L. Chen, Y. Xu, Z. Lu, J. Wu, K. Gai, P. C. K. Hung, M. Qiu, Iot microservice deployment in edge-cloud hybrid environment using reinforcement learning, IEEE Internet of Things Journal 8 (2021).
\smallskip\item \label{SP74}
Y. Yu, J. Yang, C. Guo, H. Zheng, J. He, Joint optimization of service request routing and instance placement in the microservice system, Journal of Network and Computer Applications 147 (2019).
\smallskip\item \label{SP75}
X. Zhou, X. Peng, T. Xie, J. Sun, C. Ji, D. Liu, Q. Xiang, C. He, Latent error prediction and fault localization for microservice applications by learning from system trace logs, in: Proceedings of the 2019 27th ACM Joint Meeting on European Software Engineering Conference and Symposium on the Foundations of Software Engineering, ESEC/FSE 2019, Association for Computing Machinery, New York, NY, USA, 2019.
\smallskip\item \label{SP76}
\begin{sloppypar}
I. Tsoumas, C. Symvoulidis, D. Kyriazis, Learning a generalized matrix from multi-graphs topologies towards microservices recommendations, in: K. Arai, S. Kapoor, R. Bhatia (Eds.), Intelligent Systems and Applications, Springer International Publishing, Cham, 2021.
\end{sloppypar}
\smallskip\item \label{SP77}
M. Abdullah, W. Iqbal, A. Erradi, F. Bukhari, Learning predictive autoscaling policies for cloud-hosted microservices using trace-driven modeling, in: 2019 IEEE International Conference on Cloud Computing Technology and Science (CloudCom).
\smallskip\item \label{SP78}
C. Streiffer, R. Raghavendra, T. Benson, M. Srivatsa, Learning to simplify distributed systems management, in: 2018 IEEE International Conference on Big Data (Big Data).
\smallskip\item \label{SP79}
\begin{sloppypar}
M. Imdoukh, I. Ahmad, M. G. Alfailakawi, Machine learning- based auto-scaling for containerized applications, Neural Com- puting and Applications 32 (2020).
\end{sloppypar}
\smallskip\item \label{SP80}
G. Coviello, Y. Yang, K. Rao, S. Chakradhar, Magic-pipe: Self-optimizing video analytics pipelines, in: Proceedings of the 22nd International Middleware Conference, Middleware ’21, Association for Computing Machinery, New York, NY, USA, 2021.
\smallskip\item \label{SP81}
\begin{sloppypar}
M. M. Ghorbani, F. F. Moghaddam, M. Zhang, M. Pourzandi, K. K. Nguyen, M. Cheriet, Malchain: Virtual application behaviour profiling by aggregated microservice data exchange graph, in: 2020 IEEE International Conference on Cloud Computing Technology and Science (CloudCom), 2020.
\end{sloppypar}
\smallskip\item \label{SP82}
A. Rai, R. Jagadeesh Kannan, Mathematical architecture of microservices for geographic information system based health management system, Asian Journal of Pharmaceutical and Clinical Research 10 (2017).
\smallskip\item \label{SP83}
L. Wu, J. Tordsson, J. Bogatinovski, E. Elmroth, O. Kao, Microdiag: Fine-grained performance diagnosis for microservice systems, in: 2021 IEEE/ACM International Workshop on Cloud Intelligence (CloudIntelligence).
\smallskip\item \label{SP84}
R. Oberhauser, S. Stigler, Microflows: Leveraging process mining and an automated constraint recommender for microflow modeling, in: B. Shishkov (Ed.), Business Modeling and Software Design, Springer International Publishing, Cham, 2018.
\smallskip\item \label{SP85}
D. Liu, C. He, X. Peng, F. Lin, C. Zhang, S. Gong, Z. Li, J. Ou, Z. Wu, MicroHECL: High-efficient root cause localization in large-scale microservice systems, in: 2021 IEEE/ACM 43rd International Conference on Software Engineering: Software En- gineering in Practice (ICSE-SEIP).
\smallskip\item \label{SP86}
L. Wu, J. Tordsson, E. Elmroth, O. Kao, Microrca: Root cause localization of performance issues in microservices, in: NOMS 2020 - 2020 IEEE/IFIP Network Operations and Management Symposium.
\smallskip\item \label{SP87}
G. Yu, P. Chen, Z. Zheng, Microscaler: Cost-effective scaling for microservice applications in the cloud with an online learning approach, IEEE Transactions on Cloud Computing 10 (2022).
\smallskip\item \label{SP88}
J. Yue, X. Wu, Y. Xue, Microservice aging and rejuvenation, in: 2020 World Conference on Computing and Communication Technologies (WCCCT).
\smallskip\item \label{SP89}
M. Li, D. Tang, Z. Wen, Y. Cheng, Microservice anomaly detection based on tracing data using semi-supervised learning, in: 2021 4th International Conference on Artificial Intelligence and Big Data (ICAIBD).
\smallskip\item \label{SP90}
H. Chang, M. Kodialam, T. Lakshman, S. Mukherjee, Microservice fingerprinting and classification using machine learning, in: 2019 IEEE 27th International Conference on Network Protocols (ICNP).
\smallskip\item \label{SP91}
F. Guo, B. Tang, M. Tang, H. Zhao, W. Liang, Microservice selection in edge-cloud collaborative environment: A deep reinforcement learning approach, in: 2021 8th IEEE International Conference on Cyber Security and Cloud Computing (CSCloud) /2021 7th IEEE International Conference on Edge Computing and Scalable Cloud (EdgeCom).
\smallskip\item \label{SP92}
Z. Lyu, H. Wei, X. Bai, C. Lian, Microservice-based architecture for an energy management system, IEEE Systems Journal 14 (2020).
\smallskip\item \label{SP93}
D. Rodr\'iguez-Gracia, J. A. Piedra-Fern\'andez, L. Iribarne, J. Criado, R. Ayala, J. Alonso-Montesinos, C.-U. Maria de las Mercedes, Microservices and machine learning algorithms for adaptive green buildings, Sustainability 11 (2019).
\smallskip\item \label{SP94}
F. H. Vera-Rivera, E. Puerto, H. Astudillo, C. M. Gaona, Microservices backlog–a genetic programming technique for identification and evaluation of microservices from user stories, IEEE Access 9 (2021).
\smallskip\item \label{SP95}
Z. Yang, P. Nguyen, H. Jin, K. Nahrstedt, Miras: Model-based reinforcement learning for microservice resource allocation over scientific workflows, in: 2019 IEEE 39th International Conference on Distributed Computing Systems (ICDCS).
\smallskip\item \label{SP96}
P. Nguyen, K. Nahrstedt, Monad: Self-adaptive microservice infrastructure for heterogeneous scientific workflows, in: 2017 IEEE International Conference on Autonomic Computing (ICAC), 2017.
\smallskip\item \label{SP97}
N. Parekh, S. Kurunji, A. Beck, Monitoring resources of machine learning engine in microservices architecture, in: 2018 IEEE 9th Annual Information Technology, Electronics and Mobile Communication Conference (IEMCON).
\smallskip\item \label{SP98}
S. Ji, W. Wu, Y. Pu, Multi-indicators prediction in microservice using granger causality test and attention lstm, in: 2020 IEEE World Congress on Services (SERVICES).
\smallskip\item \label{SP99}
R. Liu, P. Yang, H. Lv, W. Li, Multi-objective multi-factorial evolutionary algorithm for container placement, IEEE Transactions on Cloud Computing (2021).
\smallskip\item \label{SP100}
D. Bhamare, M. Samaka, A. Erbad, R. Jain, L. Gupta, H. A. Chan, Multi-objective scheduling of micro-services for optimal service function chains, in: 2017 IEEE International Conference on Communications (ICC).
\smallskip\item \label{SP101}
S. Horovitz, Y. Arian, M. Vaisbrot, N. Peretz, Non-intrusive cloud application transaction pattern discovery, in: 2019 IEEE 12th International Conference on Cloud Computing (CLOUD).
\smallskip\item \label{SP102}
R. Filipe, J. Correia, F. Araujo, J. Cardoso, On black-box monitoring techniques for multi-component services, in: 2018 IEEE 17th International Symposium on Network Computing and Applications (NCA).
\smallskip\item \label{SP103}
H. Alipour, Y. Liu, Online machine learning for cloud resource provisioning of microservice backend systems, in: 2017 IEEE International Conference on Big Data (Big Data).
\smallskip\item \label{SP104}
F. D. Pellegrini, F. Faticanti, M. Datar, E. Altman, D. Siracusa, Optimal blind and adaptive fog orchestration under local processor sharing, in: 2020 18th International Symposium on Modeling and Optimization in Mobile, Ad Hoc, and Wireless Networks (WiOPT).
\smallskip\item \label{SP105}
B. St \'evant, J.-L. Pazat, A. Blanc, Optimizing the performance of a microservice-based application deployed on user-provided devices, in: 2018 17th International Symposium on Parallel and Distributed Computing (ISPDC).
\smallskip\item \label{SP106}
A. Mirhosseini, S. Elnikety, T. F. Wenisch, Parslo: A gradient descent-based approach for near-optimal partial slo allotment in microservices, in: Proceedings of the ACM Symposium on Cloud Computing, SoCC ’21, Association for Computing Machinery, New York, NY, USA, 2021.
\smallskip\item \label{SP107}
D. Bajaj, U. Bharti, A. Goel, S. C. Gupta, Partial migration for re-architecting a cloud native monolithic application into microservices and faas, in: C. Badica, P. Liatsis, L. Kharb, D. Chahal (Eds.), Information, Communication and Computing Technology, Springer Singapore, Singapore, 2020.
\smallskip\item \label{SP108}
L. Wu, J. Bogatinovski, S. Nedelkoski, J. Tordsson, O. Kao, Performance diagnosis in cloud microservices using deep learning, in: H. Hacid, F. Outay, H.-y. Paik, A. Alloum, M. Petrocchi, M. R. Bouadjenek, A. Beheshti, X. Liu, A. Maaradji (Eds.), Service-Oriented Computing – ICSOC 2020 Workshops, Springer International Publishing, Cham, 2021.
\smallskip\item \label{SP109}
B. Choi, J. Park, C. Lee, D. Han, Phpa: A proactive autoscaling framework for microservice chain, in: 5th Asia-Pacific Workshop on Networking (APNet 2021), APNet 2021, Association for Computing Machinery, New York, NY, USA, 2022.
\smallskip\item \label{SP110}
J. Rahman, P. Lama, Predicting the End-to-End Tail Latency of Containerized Microservices in the Cloud, in: 2019 IEEE International Conference on Cloud Engineering (IC2E).
\smallskip\item \label{SP111}
M. Abdullah, W. Iqbal, A. Mahmood, F. Bukhari, A. Erradi, Predictive autoscaling of microservices hosted in fog microdata center, IEEE Systems Journal 15 (2021).
\smallskip\item \label{SP112}
N. Marie-Magdelaine, T. Ahmed, Proactive autoscaling for cloud-native applications using machine learning, in: GLOBECOM 2020 - 2020 IEEE Global Communications Conference.
\smallskip\item \label{SP113}
K. Ray, A. Banerjee, N. C. Narendra, Proactive microservice placement and migration for mobile edge computing, in: 2020 IEEE/ACM Symposium on Edge Computing (SEC).
\smallskip\item \label{SP114}
J. Bender, J. Ovtcharova, Prototyping machine-learning-supported lead time prediction using automl, Procedia Computer Science 180 (2021) 649–655. Proceedings of the 2nd International Conference on Industry 4.0 and Smart Manufacturing (ISM 2020).
\smallskip\item \label{SP115}
Q. Li, B. Li, P. Mercati, R. Illikkal, C. Tai, M. Kishinevsky, C. Kozyrakis, Rambo: Resource allocation for microservices using bayesian optimization, IEEE Computer Architecture Letters 20 (2021).
\smallskip\item \label{SP116}
A. Belhadi, Y. Djenouri, G. Srivastava, J. C.-W. Lin, Reinforcement learning multi-agent system for faults diagnosis of mircoservices in industrial settings, Computer Communications 177 (2021).
\smallskip\item \label{SP118}
P. Li, J. Song, H. Xu, L. Dong, Y. Zhou, Resource scheduling optimisation algorithm for containerised microservice architecture in cloud computing, International Journal of High Performance Systems Architecture 8 (2018).
\smallskip\item \label{SP119}
P. Kang, P. Lama, Robust resource scaling of containerized microservices with probabilistic machine learning, in: 2020 IEEE/ACM 13th International Conference on Utility and Cloud Computing (UCC).
\smallskip\item \label{SP120}
A. Bali, M. Al-Osta, S. Ben Dahsen, A. Gherbi, Rule based auto-scalability of iot services for efficient edge device resource utilization, Journal of Ambient Intelligence and Humanized Computing 11 (2020).
\smallskip\item \label{SP121}
Y. Gan, M. Liang, S. Dev, D. Lo, C. Delimitrou, Sage: Practical and scalable ml-driven performance debugging in microservices, in: Proceedings of the 26th ACM International Conference on Architectural Support for Programming Languages and Operating Systems, ASPLOS ’21, Association for Computing Machinery, New York, NY, USA, 2021.
\smallskip\item \label{SP122}
S.-P. Ma, Y. Chuang, C.-W. Lan, H.-M. Chen, C.-Y. Huang, C.-Y. Li, Scenario-based microservice retrieval using word2vec, in: 2018 IEEE 15th International Conference on e-Business Engineering (ICEBE), 2018.
\smallskip\item \label{SP123}
H. Gu, X. Li, M. Liu, S. Wang, Scheduling method with adaptive learning for microservice workflows with hybrid resource provisioning, International Journal of Machine Learning and Cybernetics 12 (2021).
\smallskip\item \label{SP124}
Y. Gan, Y. Zhang, K. Hu, D. Cheng, Y. He, M. Pancholi, C. Delimitrou, Seer: Leveraging big data to navigate the complexity of performance debugging in cloud microservices, in: Proceedings of the Twenty-Fourth International Conference on Architectural Support for Programming Languages and Operating Systems, ASPLOS ’19, Association for Computing Machinery, New York, NY, USA, 2019.
\smallskip\item \label{SP125}
\begin{sloppypar}
F. Rossi, V. Cardellini, F. L. Presti, Self-adaptive threshold-based policy for microservices elasticity, in: 2020 28th International Symposium on Modeling, Analysis, and Simulation of Computer and Telecommunication Systems (MASCOTS).
\end{sloppypar}
\smallskip\item \label{SP126}
J. Bogatinovski, S. Nedelkoski, J. Cardoso, O. Kao, Self-supervised anomaly detection from distributed traces, in: 2020 IEEE/ACM 13th International Conference on Utility and Cloud Computing (UCC), 2020.
\smallskip\item \label{SP127}
D. Petcu, Service deployment challenges in cloud-to-edge continuum, Scalable Computing: Practice and Experience 22 (2021).
\smallskip\item \label{SP128}
P. Kr\"amer, P. Diederich, C. Kr\"amer, R. Pries, W. Kellerer, A. Blenk, sfc2cpu: Operating a service function chain platform with neural combinatorial optimization, in: 2021 IFIP/IEEE International Symposium on Integrated Network Management (IM).
\smallskip\item \label{SP129}
Y. Li, T. Li, P. Shen, L. Hao, W. Liu, S. Wang, Y. Song, L. Bao, Sim-drs: a similarity-based dynamic resource scheduling algorithm for microservice-based web systems, PeerJ Computer Science 7 (2021).
\smallskip\item \label{SP130}
Y. Zhang, W. Hua, Z. Zhou, G. E. Suh, C. Delimitrou, Sinan: Ml-based and qos-aware resource management for cloud microservices, in: Proceedings of the 26th ACM International Conference on Architectural Support for Programming Languages and Operating Systems, ASPLOS ’21, Association for Computing Machinery, New York, NY, USA, 2021.
\smallskip\item \label{SP131}
J. Herrera, G. Molt \'o, Toward bio-inspired auto-scaling algorithms: An elasticity approach for container orchestration platforms, IEEE Access 8 (2020).
\smallskip\item \label{SP132}
P. Zhao, P. Wang, X. Yang, J. Lin, Towards cost-efficient edge intelligent computing with elastic deployment of container-based microservices, IEEE Access 8 (2020).
\smallskip\item \label{SP133}
G. Somashekar, A. Gandhi, Towards optimal configuration of microservices, in: Proceedings of the 1st Workshop on Machine Learning and Systems, EuroMLSys ’21, Association for Computing Machinery, New York, NY, USA, 2021.
\smallskip\item \label{SP134}
H. Chen, K. Wei, A. Li, T. Wang, W. Zhang, Trace-based intelligent fault diagnosis for microservices with deep learning, in: 2021 IEEE 45th Annual Computers, Software, and Applications Conference (COMPSAC).
\smallskip\item \label{SP135}
P. Liu, H. Xu, Q. Ouyang, R. Jiao, Z. Chen, S. Zhang, J.Yang, L. Mo, J. Zeng, W. Xue, D. Pei, Unsupervised detection of microservice trace anomalies through service-level deep bayesian networks, in: 2020 IEEE 31st International Symposium on Software Reliability Engineering (ISSRE), 2020.
\smallskip\item \label{SP136}
M. Abdullah, W. Iqbal, A. Erradi, Unsupervised learning approach for web application auto-decomposition into microservices, Journal of Systems and Software 151 (2019).
\smallskip\item \label{SP137}
W. Liang, Y. Hu, X. Zhou, Y. Pan, K. I.-K. Wang, Variational few-shot learning for microservice-oriented intrusion detection in distributed industrial iot, IEEE Transactions on Industrial Informatics 18 (2022).
\smallskip\item \label{SP138}
H. Sami, A. Mourad, W. El-Hajj, Vehicular-obus-as-on-demand-fogs: Resource and context aware deployment of containerized micro-services, IEEE/ACM Transactions on Networking 28 (2020).
\smallskip\item \label{SP139}
P. Petrou, S. Karagiorgou, D. Alexandrou, Weighted load balancing mechanisms over streaming big data for online machine learning, in: EDBT/ICDT Workshops, 2021.
\smallskip\item \label{SP140} H. Bangui, B. Rossi, B. Buhnova, A Conceptual Antifragile Microservice Framework for Reshaping Critical Infrastructures, in: 2022 IEEE International Conference on Software Maintenance and Evolution (ICSME), 2022.
\smallskip\item \label{SP141} A. Ali, M. M. Iqbal, A Cost and Energy Efficient Task Scheduling Technique to Offload Microservices Based Applications in Mobile Cloud Computing, IEEE Access 10 (2022).
\smallskip\item \label{SP142} Y. Qi, S. Shao, S. Wu, X. Qiu, S. Guo, S. Guo, A Distributed Intelligent Service Trusted Provision Approach for IoT, IEEE Internet of Things Journal 10 (2023).
\smallskip\item \label{SP143} H. Li, Y. Zhao, Z. Liu, W. Liu, A Distributed Microservice Scheduling Optimization Method, in: 2023 IEEE International Conference on Control, Electronics and Computer Technology (ICCECT), 2023.
\smallskip\item \label{SP144} Bagehorn F., Rios J., Jha S., Filepp R., Shwartz L., Abe N., Yang X., A fault injection platform for learning AIOps models, in: ACM International Conference Proceeding Series, 2022.
\smallskip\item \label{SP145} Hossein Ebrahimpour, Mehrdad Ashtiani, Fatemeh Bakhshi, Ghazaleh Bakhtiariazad, A heuristic-based package-aware function scheduling approach for creating a trade-off between cold start time and cost in FaaS computing environments, The Journal of Supercomputing 79 (2023).
\smallskip\item \label{SP147} Thiago Felipe da Silva Pinheiro, Paulo Pereira, Bruno Silva, Paulo Maciel, A performance modeling framework for microservices-based cloud infrastructures, The Journal of Supercomputing 79 (2023).
\smallskip\item \label{SP148} A. Heimerson, J. Eker, K. -E. Årzén, A Proactive Cloud Application Auto-Scaler using Reinforcement Learning, in: 2022 IEEE/ACM 15th International Conference on Utility and Cloud Computing (UCC), 2022.
\smallskip\item \label{SP149} C. H. Zhang, M. Omair Shafiq, A Real-time, Scalable Monitoring and User Analytics Solution for Microservices-based Software Applications, in: 2022 IEEE International Conference on Big Data (Big Data), 2022.
\smallskip\item \label{SP150} Sedigheh Khoshnevis, A search-based identification of variable microservices for enterprise SaaS, Frontiers of Computer Science (2022).
\smallskip\item \label{SP151} D. Zhou, H. Chen, G. Cheng, A Security Containers Placement Algorithm Based on DQN for Microservices to Defend Against Co-Resident Threat, in: 2023 8th International Conference on Computer and Communication Systems (ICCCS), 2023.
\smallskip\item \label{SP152} C. -A. Sun, T. Zeng, W. Zuo, H. Liu, A Trace-Log-Clusterings-Based Fault Localization Approach to Microservice Systems, in: 2023 IEEE International Conference on Web Services (ICWS), 2023.
\smallskip\item \label{SP153} Claudia Canali, Giuseppe Di Modica, Riccardo Lancellotti, Stefano Rossi, Domenico Scotece, A Validated Performance Model for Micro-services Placement in Fog Systems, SN Computer Science 4 (2023).
\smallskip\item \label{SP154} C. Duan, T. Jia, Y. Li, G. Huang, AcLog: An Approach to Detecting Anomalies from System Logs with Active Learning, in: 2023 IEEE International Conference on Web Services (ICWS), 2023.
\smallskip\item \label{SP155} Ruibo Chen, Yanjun Pu, Bowen Shi, Wenjun Wu, An automatic model management system and its implementation for AIOps on microservice platforms, The Journal of Supercomputing 79 (2023).
\smallskip\item \label{SP156} Mansoureh Zare, Yasser Elmi Sola, Hesam Hasanpour, An autonomous planning model for solving IoT service placement problem using the imperialist competitive algorithm, The Journal of Supercomputing 79 (2023).
\smallskip\item \label{SP157} Y. Sever, G. Ekinci, A. H. Dogan, B. Alparslan, A. S. Gurbuz, V. Jabrayilov, P. Angin, An Empirical Analysis of IDS Approaches in Container Security, in: 2022 International Workshop on Secure and Reliable Microservices and Containers (SRMC), 2022.
\smallskip\item \label{SP158} Wenliang LinYilie HeZhongliang DengKe WangBin JinXiaotian Zhou, An end-to-end software-defined network framework and optimal service development model for SAGN, Telecommunication Systems (2022).
\smallskip\item \label{SP159} Wesley K. G. Assunção, Thelma Elita Colanzi, Luiz Carvalho, Alessandro Garcia, Juliana Alves Pereira, Maria Julia de Lima, Carlos Lucena, Analysis of a many-objective optimization approach for identifying microservices from legacy systems, Empirical Software Engineering 27 (2022).
\smallskip\item \label{SP160} Iman Kohyarnejadfard, Daniel Aloise, Seyed Vahid Azhari, Michel R. Dagenais, Anomaly detection in microservice environments using distributed tracing data analysis and NLP, Journal of Cloud Computing 11 (2022).
\smallskip\item \label{SP161} M. Sowmya, A. J. Rai, V. Spoorthi, M. Irfan, P. B. Honnavalli, S. Nagasundari, API Traffic Anomaly Detection in Microservice Architecture, in: 2023 IEEE/ACM 23rd International Symposium on Cluster, Cloud and Internet Computing Workshops (CCGridW), 2023.
\smallskip\item \label{SP162} L. S. Hettiarachchi, S. V. Jayadeva, R. A. V. Bandara, D. Palliyaguruge, U. S. S. S. Arachchillage, D. Kasthurirathna, Artificial Intelligence-Based Centralized Resource Management Application for Distributed Systems, in: 2022 13th International Conference on Computing Communication and Networking Technologies (ICCCNT), 2022.
\smallskip\item \label{SP163} Rafael de Jesus Martins, Juliano Araújo Wickboldt, Lisandro Zambenedetti Granville, Assisted Monitoring and Security Provisioning for 5G Microservices-Based Network Slices with SWEETEN, Journal of Network and Systems Management 31 (2023).
\smallskip\item \label{SP164} S. Choochotkaew, T. Chiba, S. Trent, T. Yoshimura, M. Amaral, AutoDECK: Automated Declarative Performance Evaluation and Tuning Framework on Kubernetes, in: 2022 IEEE 15th International Conference on Cloud Computing (CLOUD), 2022.
\smallskip\item \label{SP165} T. Miyazawa, M. Jibiki, V. P. Kafle, Automated Data Analytics and Resource Arbitration Scheduling for Containerized Network Functions, in: 2022 IEEE Future Networks World Forum (FNWF), 2022.
\smallskip\item \label{SP166} A. A. Pramesti, A. I. Kistijantoro, Autoscaling Based on Response Time Prediction for Microservice Application in Kubernetes, in: 2022 9th International Conference on Advanced Informatics: Concepts, Theory and Applications (ICAICTA), 2022.
\smallskip\item \label{SP167} Ahmet Vedat TokmakAkhan AkbulutCagatay Catal, Boosting the visibility of services in microservice architecture, Cluster Computing (2023).
\smallskip\item \label{SP168} A. Boukhtouta, T. Madi, M. Pourzandi, H. A. A., Cloud Native Applications Profiling using a Graph Neural Networks Approach, in: 2022 IEEE Future Networks World Forum (FNWF), 2022.
\smallskip\item \label{SP169} M. D. Hossain, T. Sultana, S. Akhter, M. I. Hossain, G. -W. Lee, C. S. Hong, E. -N. Huh, Computation Offloading Strategy Based on Multi-armed Bandit Learning in Microservice-enabled Vehicular Edge Computing Networks, in: 2023 International Conference on Information Networking (ICOIN), 2023.
\smallskip\item \label{SP170} D. Ou, C. Jiang, M. Zheng, Y. Ren, Container Power Consumption Prediction Based on GBRT-PL for Edge Servers in Smart City, IEEE Internet of Things Journal 10 (2023).
\smallskip\item \label{SP171} Z. Wang, C. -A. Sun, M. Aiello, Context-aware IoT Service Recommendation: A Deep Collaborative Filtering-based Approach, in: 2022 IEEE International Conference on Web Services (ICWS), 2022.
\smallskip\item \label{SP172} Juan Luis Herrera, Javier Berrocal, Stefano Forti, Antonio Brogi, Juan M. Murillo, Continuous QoS-aware adaptation of Cloud-IoT application placements, Computing 105 (2023).
\smallskip\item \label{SP173} M. Xu, C. Song, S. Ilager, S. S. Gill, J. Zhao, K. Ye, C. Xu, CoScal: Multifaceted Scaling of Microservices With Reinforcement Learning, IEEE Transactions on Network and Service Management 19 (2022).
\smallskip\item \label{SP174} Abdullah LakhanMuhammad Suleman MemonQurat-ul-ain MastoiMohamed ElhosenyMazin Abed MohammedMumtaz QabulioMohamed Abdel-Basset, Cost-efficient mobility offloading and task scheduling for microservices IoVT applications in container-based fog cloud network, Cluster Computing (2022).
\smallskip\item \label{SP175} P. Krämer, P. Diederich, C. Krämer, R. Pries, W. Kellerer, A. Blenk, D2A: Operating a Service Function Chain Platform With Data-Driven Scheduling Policies, IEEE Transactions on Network and Service Management 19 (2022).
\smallskip\item \label{SP176} Abdullah Alelyani, Amitava Datta, Ghulam Mubashar Hassan, DAScheduler: Dependency-Aware Scheduling Algorithm for Containerized Dependent Jobs, Journal of Grid Computing 21 (2023).
\smallskip\item \label{SP177} P. Benedetti, G. Coviello, K. Rao, S. Chakradhar, DataX Allocator: Dynamic resource management for stream analytics at the Edge, in: 2022 9th International Conference on Internet of Things: Systems, Management and Security (IOTSMS), 2022.
\smallskip\item \label{SP178} Y. Chen, M. Yan, D. Yang, X. Zhang, Z. Wang, Deep Attentive Anomaly Detection for Microservice Systems with Multimodal Time-Series Data, in: 2022 IEEE International Conference on Web Services (ICWS), 2022.
\smallskip\item \label{SP179} Z. Guo, K. Yu, Z. Lv, K. -K. R. Choo, P. Shi, J. J. P. C. Rodrigues, Deep Federated Learning Enhanced Secure POI Microservices for Cyber-Physical Systems, IEEE Wireless Communications 29 (2022).
\smallskip\item \label{SP180} C. Wang, B. Jia, H. Yu, X. Li, X. Wang, T. Taleb, Deep Reinforcement Learning for Dependency-aware Microservice Deployment in Edge Computing, in: GLOBECOM 2022 - 2022 IEEE Global Communications Conference, 2022.
\smallskip\item \label{SP181} Feiyan Guo, Bing Tang, Mingdong Tang, Wei Liang, Deep reinforcement learning-based microservice selection in mobile edge computing, Cluster Computing 26 (2023).
\smallskip\item \label{SP182} Q. Zhu, S. Wang, H. Huang, Y. Lei, W. Zhan, H. Duan, Deep-Reinforcement-Learning-Based Service Placement for Video Analysis in Edge Computing, in: 2023 8th International Conference on Cloud Computing and Big Data Analytics (ICCCBDA), 2023.
\smallskip\item \label{SP183} C. Zhang, X. Peng, C. Sha, K. Zhang, Z. Fu, X. Wu, Q. Lin, D. Zhang, DeepTraLog: Trace-Log Combined Microservice Anomaly Detection through Graph-based Deep Learning, in: 2022 IEEE/ACM 44th International Conference on Software Engineering (ICSE), 2022.
\smallskip\item \label{SP184} L. Traini, V. Cortellessa, DeLag: Using Multi-Objective Optimization to Enhance the Detection of Latency Degradation Patterns in Service-Based Systems, IEEE Transactions on Software Engineering (2023).
\smallskip\item \label{SP185} X. Yu, W. Wu, Y. Wang, Dependable Workflow Scheduling for Microservice QoS Based on Deep Q-Network, in: 2022 IEEE International Conference on Web Services (ICWS), 2022.
\smallskip\item \label{SP186} Lizhe ChenJi WuHaiyan YangKui Zhang, Does PageRank apply to service ranking in microservice regression testing?, Software Quality Journal (2022).
\smallskip\item \label{SP187} Sheuli Chakraborty, Debashis De, Kaushik Mazumdar, DoME: Dew computing based microservice execution in mobile edge using Q-learning, Applied Intelligence 53 (2023).
\smallskip\item \label{SP188} Z. Xiao, S. Hu, DScaler: A Horizontal Autoscaler of Microservice Based on Deep Reinforcement Learning, in: 2022 23rd Asia-Pacific Network Operations and Management Symposium (APNOMS), 2022.
\smallskip\item \label{SP189} Saravanan Muniswamy, Radhakrishnan Vignesh, DSTS: A hybrid optimal and deep learning for dynamic scalable task scheduling on container cloud environment, Journal of Cloud Computing 11 (2022).
\smallskip\item \label{SP190} S. B. Chetty, H. Ahmadi, M. Tornatore, A. Nag, Dynamic Decomposition of Service Function Chain Using a Deep Reinforcement Learning Approach, IEEE Access 10 (2022).
\smallskip\item \label{SP191} F. Rossi, V. Cardellini, F. L. Presti, M. Nardelli, Dynamic Multi-Metric Thresholds for Scaling Applications Using Reinforcement Learning, IEEE Transactions on Cloud Computing 11 (2023).
\smallskip\item \label{SP192} C. Lee, T. Yang, Z. Chen, Y. Su, M. R. Lyu, Eadro: An End-to-End Troubleshooting Framework for Microservices on Multi-source Data, in: 2023 IEEE/ACM 45th International Conference on Software Engineering (ICSE), 2023.
\smallskip\item \label{SP193} J. Qi, H. Zhang, X. Li, H. Ji, X. Shao, Edge-edge Collaboration Based Micro-service Deployment in Edge Computing Networks, in: 2023 IEEE Wireless Communications and Networking Conference (WCNC), 2023.
\smallskip\item \label{SP194} I. Syrigos, D. Kefalas, N. Makris, T. Korakis, EELAS: Energy Efficient and Latency Aware Scheduling of Cloud-Native ML Workloads, in: 2023 15th International Conference on COMmunication Systems \& NETworkS (COMSNETS), 2023.
\smallskip\item \label{SP195} Mohamed Hedi Fourati, Soumaya Marzouk, Mohamed Jmaiel, EPMA: Elastic Platform for Microservices-based Applications: Towards Optimal Resource Elasticity, Journal of Grid Computing 20 (2022).
\smallskip\item \label{SP196} L. S. Hettiarachchi, S. V. Jayadeva, R. A. V. Bandara, D. Palliyaguruge, U. S. S. S. Arachchillage, D. Kasthurirathna, Expert System for Kubernetes Cluster Autoscaling and Resource Management, in: 2022 4th International Conference on Advancements in Computing (ICAC), 2022.
\smallskip\item \label{SP197} Mohammad Hadi Dehghani, Shekoufeh Kolahdouz-Rahimi, Massimo Tisi, Dalila Tamzalit, Facilitating the migration to the microservice architecture via model-driven reverse engineering and reinforcement learning, Software and Systems Modeling 21 (2022).
\smallskip\item \label{SP198} Kawasaki J., Koyama D., Miyasaka T., Otani T., Failure Prediction in Cloud Native 5G Core With eBPF-based Observability, in: IEEE Vehicular Technology Conference, 2023.
\smallskip\item \label{SP199} Wang Q., Rios J., Jha S., Shanmugam K., Bagehorn F., Yang X., Filepp R., Abe N., Shwartz L., Fault Injection Based Interventional Causal Learning for Distributed Applications, in: Proceedings of the 37th AAAI Conference on Artificial Intelligence, AAAI 2023, 2023.
\smallskip\item \label{SP200} X. Yang, J. Wang, B. Zhou, W. Wang, W. Liu, Y. Dong, Fine-grained Spatiotemporal Features-Based for Anomaly Detection in Microservice Systems, in: 2022 IEEE 24th Int Conf on High Performance Computing \& Communications, 8th Int Conf on Data Science \& Systems, 20th Int Conf on Smart City, 8th Int Conf on Dependability in Sensor, Cloud \& Big Data Systems \& Application (HPCC/DSS/SmartCity/DependSys), 2022.
\smallskip\item \label{SP201} R. Ren, Y. Wang, F. Liu, Z. Li, G. Tyson, T. Miao, G. Xie, Grace: Interpretable Root Cause Analysis by Graph Convolutional Network for Microservices, in: 2023 IEEE/ACM 31st International Symposium on Quality of Service (IWQoS), 2023.
\smallskip\item \label{SP202} W. Lv, P. Yang, T. Zheng, C. Lin, Z. Wang, M. Deng, Q. Wang, Graph Reinforcement Learning-based Dependency-Aware Microservice Deployment in Edge Computing, in: IEEE Internet of Things Journal, 2023.
\smallskip\item \label{SP203} H. X. Nguyen, S. Zhu, M. Liu, Graph-PHPA: Graph-based Proactive Horizontal Pod Autoscaling for Microservices using LSTM-GNN, in: 2022 IEEE 11th International Conference on Cloud Networking (CloudNet), 2022.
\smallskip\item \label{SP204} H. He, L. Su, K. Ye, GraphGRU: A Graph Neural Network Model for Resource Prediction in Microservice Cluster, in: 2022 IEEE 28th International Conference on Parallel and Distributed Systems (ICPADS), 2023.
\smallskip\item \label{SP205} T. Rathod, C. T. Joseph, J. P. Martin, Improving Industry 4.0 Readiness: Monolith Application Refactoring using Graph Attention Networks, in: 2023 IEEE/ACM 23rd International Symposium on Cluster, Cloud and Internet Computing Workshops (CCGridW), 2023.
\smallskip\item \label{SP206} Abeer Abdel Khaleq, Ilkyeun Ra, Intelligent microservices autoscaling module using reinforcement learning, Cluster Computing 26 (2023).
\smallskip\item \label{SP207} Y. Zhang, C. Li, N. Chen, P. Zhang, Intelligent Requests Orchestration for Microservice Management Based on Blockchain in Software Defined Networking: a Security Guarantee, in: 2022 IEEE International Conference on Communications Workshops (ICC Workshops), 2022.
\smallskip\item \label{SP208} Thijs Metsch, Magdalena Viktorsson, Adrian Hoban, Monica Vitali, Ravi Iyer, Erik Elmroth, Intent-Driven Orchestration: Enforcing Service Level Objectives for Cloud Native Deployments, SN Computer Science 4 (2023).
\smallskip\item \label{SP209} G. Pearce, A. Pflaum, D. A. Balasoiu, C. Szabo, Jeopardy Assessment for Dynamic Configuration of Collaborative Microservice Architectures, in: 2022 Winter Simulation Conference (WSC), 2022.
\smallskip\item \label{SP210} Feiyan Guo, Bing Tang, Mingdong Tang, Joint optimization of delay and cost for microservice composition in mobile edge computing, World Wide Web 25 (2022).
\smallskip\item \label{SP211} Javad Dogani, Farshad Khunjush, Mehdi Seydali, K-AGRUED: A Container Autoscaling Technique for Cloud-based Web Applications in Kubernetes Using Attention-based GRU Encoder-Decoder, Journal of Grid Computing 20 (2022).
\smallskip\item \label{SP212} S. Hirai, H. Baba, M. Matsumoto, T. Hamano, K. Noguchi, Machine Learning based Performance Prediction for Cloud-native 5G Mobile Core Network, in: 2022 IEEE Wireless Communications and Networking Conference (WCNC), 2022.
\smallskip\item \label{SP213} Yang L., Li J., Shi K., Yang S., Yang Q., Sun J., MicroMILTS: Fault Location for Microservices Based Mutual Information and LSTM Autoencoder, in: APNOMS 2022 - 23rd Asia-Pacific Network Operations and Management Symposium: Data-Driven Intelligent Management in the Era of beyond 5G, 2022.
\smallskip\item \label{SP214} W. Lv, Q. Wang, P. Yang, Y. Ding, B. Yi, Z. Wang, C. Lin, Microservice Deployment in Edge Computing Based on Deep Q Learning, IEEE Transactions on Parallel and Distributed Systems 33 (2022).
\smallskip\item \label{SP215} Y. Liu, B. Yang, X. Yang, Y. Wu, C. Li, Microservice Dynamic Migration based on Age of Service for Edge Computing, in: 2022 IEEE International Conference on Industrial Technology (ICIT), 2022.
\smallskip\item \label{SP216} W. Cruz, L. D. Michel, B. Drozdenko, S. Roodbeen, ML and Network Traces to M.A.R.S, in: 2023 IEEE International Conference on Cyber Security and Resilience (CSR), 2023.
\smallskip\item \label{SP217} K. Sooksatra, R. Maharjan, T. Cerny, Monolith to Microservices: VAE-Based GNN Approach with Duplication Consideration, in: 2022 IEEE International Conference on Service-Oriented System Engineering (SOSE), 2022.
\smallskip\item \label{SP218} Z. Li, H. Sun, Z. Xiong, Q. Huang, Z. Hu, D. Li, S. Ruan, H. Hong, J. Gui, J. He, Z. Xu, Y. Fang, Noah: Reinforcement-Learning-Based Rate Limiter for Microservices in Large-Scale E-Commerce Services, IEEE Transactions on Neural Networks and Learning Systems 34 (2023).
\smallskip\item \label{SP219} Y. Yu, J. Liu, J. Fang, Online Microservice Orchestration for IoT via Multiobjective Deep Reinforcement Learning, IEEE Internet of Things Journal 9 (2022).
\smallskip\item \label{SP220} A. Hrusto, E. Engström, P. Runeson, Optimization of Anomaly Detection in a Microservice System Through Continuous Feedback from Development, in: 2022 IEEE/ACM 10th International Workshop on Software Engineering for Systems-of-Systems and Software Ecosystems (SESoS), 2022.
\smallskip\item \label{SP221} X. Chen, Y. Wu, S. Xiao, Particle Swarm–Grey Wolf Cooperation Algorithm Based on Microservice Container Scheduling Problem, IEEE Access 11 (2023).
\smallskip\item \label{SP222} Y. Gan, M. Liang, S. Dev, D. Lo, C. Delimitrou, Practical and Scalable ML-Driven Cloud Performance Debugging With Sage, IEEE Micro 42 (2022).
\smallskip\item \label{SP223} Al Qassem L.M., Stouraitis T., Damiani E., Elfadel I.A.M., Proactive Random-Forest Autoscaler for Microservice Resource Allocation, IEEE Access 11 (2023).
\smallskip\item \label{SP224} B. Jeong, J. Jeon, Y. -S. Jeong, Proactive Resource Autoscaling Scheme based on SCINet for High-performance Cloud Computing, IEEE Transactions on Cloud Computing 11 (2023).
\smallskip\item \label{SP225} K. Zhang, C. Zhang, X. Peng, C. Sha, PUTraceAD: Trace Anomaly Detection with Partial Labels based on GNN and PU Learning, in: 2022 IEEE 33rd International Symposium on Software Reliability Engineering (ISSRE), 2022.
\smallskip\item \label{SP226} K. R. Sheshadri, J. Lakshmi, QoS aware FaaS for Heterogeneous Edge-Cloud continuum, in: 2022 IEEE 15th International Conference on Cloud Computing (CLOUD), 2022.
\smallskip\item \label{SP227} G. Somashekar, A. Suresh, S. Tyagi, V. Dhyani, K. Donkada, A. Pradhan, A. Gandhi, Reducing the Tail Latency of Microservices Applications via Optimal Configuration Tuning, in: 2022 IEEE International Conference on Autonomic Computing and Self-Organizing Systems (ACSOS), 2022.
\smallskip\item \label{SP228} J. E. Joyce, S. Sebastian, Reinforcement Learning based Autoscaling for Kafka-centric Microservices in Kubernetes, in: 2022 IEEE 4th PhD Colloquium on Emerging Domain Innovation and Technology for Society (PhD EDITS), 2022.
\smallskip\item \label{SP229} J. Zhao, C. Su, Y. Wang, Research on Microservice Coordination Technologies based on Deep Reinforcement Learning, in: 2022 2nd International Conference on Electronic Information Technology and Smart Agriculture (ICEITSA), 2022.
\smallskip\item \label{SP230} Park J., Son J., Kim D., Resource Metric Refining Module for AIOps Learning Data in Kubernetes Microservice, KSII Transactions on Internet and Information Systems 17 (2023).
\smallskip\item \label{SP231} S. M. Rajagopal, M. Supriya, R. Buyya, Resource Provisioning Using Meta-Heuristic Methods for IoT Microservices With Mobility Management, IEEE Access 11 (2023).
\smallskip\item \label{SP232} Zhengzhe XiangYuhang ZhengDongjing WangMengzhu HeCheng ZhangZengwei Zheng, Robust and Cost-effective Resource Allocation for Complex IoT Applications in Edge-Cloud Collaboration, Mobile Networks and Applications (2022).
\smallskip\item \label{SP233} S. Zhang, P. Jin, Z. Lin, Y. Sun, B. Zhang, S. Xia, Z. Li, Z. Zhong, M. Ma, W. Jin, D. Zhang, Z. Zhu, D. Pei, Robust Failure Diagnosis of Microservice System through Multimodal Data, IEEE Transactions on Services Computing 6 (2023).
\smallskip\item \label{SP234} S. P. Kadiyala, X. Li, W. Lee, A. Catlin, Securing Microservices Against Password Guess Attacks using Hardware Performance Counters, in: 2022 IEEE 35th International System-on-Chip Conference (SOCC), 2022.
\smallskip\item \label{SP235} T. Stojanovic, S. D. Lazarevi\'c, The Application of ChatGPT for Identification of Microservices, in: E-business technologies conference proceedings, 2023.
\smallskip\item \label{SP236} H. Zeng, T. Wang, A. Li, Y. Wu, H. Wu, W. Zhang, Topology-Aware Self-Adaptive Resource Provisioning for Microservices, in: 2023 IEEE International Conference on Web Services (ICWS), 2023.
\smallskip\item \label{SP237} Zeb S., Rathore M.A., Hassan S.A., Raza S., Dev K., Fortino G., Toward AI-Enabled NextG Networks with Edge Intelligence-Assisted Microservice Orchestration, IEEE Wireless Communications 30 (2023).
\smallskip\item \label{SP238} R. Ren, Y. Wang, F. Liu, Z. Li, G. Xie, Triple:The Interpretable Deep Learning Anomaly Detection Framework based on Trace-Metric-Log of Microservice, in: 2023 IEEE/ACM 31st International Symposium on Quality of Service (IWQoS), 2023.
\smallskip\item \label{SP239} F. Dressler, C. F. Chiasserini, F. H. P. Fitzek, H. Karl, R. L. Cigno, A. Capone, C. Casetti, F. Malandrino, V. Mancuso, F. Klingler, G. Rizzo, V-Edge: Virtual Edge Computing as an Enabler for Novel Microservices and Cooperative Computing, IEEE Network 36 (2022).
\smallskip\item \label{SP240} Mekki M., Brik B., Ksentini A., Verikoukis C., XAI-Enabled Fine Granular Vertical Resources Autoscaler, in: 2023 IEEE 9th International Conference on Network Softwarization: Boosting Future Networks through Advanced Softwarization, NetSoft 2023 - Proceedings, 2023.
\smallskip\item \label{SP241} Ruibo Chen, Jian Ren, Lingfeng Wang, Yanjun Pu, Kaiyuan Yang \& Wenjun Wu, MicroEGRCL: An Edge-Attention-Based Graph Neural Network Approach for Root Cause Localization in Microservice Systems, in: Service-Oriented Computing. ICSOC 2022. Lecture Notes in Computer Science, vol 13740., 2022.
\smallskip\item \label{SP242} Lingzhi Wang, Nengwen Zhao, Junjie Chen, Pinnong Li, Wenchi Zhang, Kaixin Sui, Root-Cause Metric Location for Microservice Systems via Log Anomaly Detection, in: 2020 IEEE International Conference on Web Services (ICWS), 2020.
\smallskip\item \label{SP243} Zhijun Ding, Song Wang, Changjun Jiang, Kubernetes-Oriented Microservice Placement With Dynamic Resource Allocation, IEEE Transactions on Cloud Computing (2022).
\smallskip\item \label{SP244} Sasho Nedelkoski, Jorge Cardoso, Odej Kao, Anomaly Detection from System Tracing Data Using Multimodal Deep Learning, in: IEEE Cloud, 2019.
\smallskip\item \label{SP245} Rajsimman Ravichandiran, Hadi Bannazadeh, Alberto Leon-Garcia, Anomaly Detection using Resource Behaviour Analysis for Autoscaling systems, in: NetSoft, 2018.
\smallskip\item \label{SP246} Mohammad Javad Amiri, Object-Aware Identification of Microservices, in: IEEE SCC, 2018.
\smallskip\item \label{SP247} Yukun Zhang, Bo Liu, Liyun Dai, Kang Chen, Xuelian Cao, Automated Microservice Identification in Legacy Systems with Functional and Non-Functional Metrics, in: IEEE ICSA, 2020.
\smallskip\item \label{SP248} Sinan Eski, Feza Buzluca, An automatic extraction approach: transition to microservices architecture from monolithic application, in: XP 2018, 2018.
\smallskip\item \label{SP249} Luiz Carvalho Alessandro Garcia, Thelma Elita Colanzi, Wesley K. G. Assunção, Maria Julia Lima, Baldoino Fonseca Márcio Ribeiro, Carlos Lucena, Search-based many-criteria identification of microservices from legacy systems., in: GECCO, 2020.
\smallskip\item \label{SP250} Mohamed Daoud, Asmae El Mezouari, Noura Faci, Djamal Benslimane, Zakaria Maamar \& Aziz El Fazziki, Automatic Microservices Identification from a Set of Business Processes, in: SADASC, 2020.
\smallskip\item \label{SP251} Malak Saidi, Anis Tissaoui, Sami Faiz, A DDD Approach Towards Automatic Migration To Microservices, in: ASET, 2023.
\smallskip\item \label{SP252} Chenghao Song, Minxian Xu, Kejiang Ye, Huaming Wu, Sukhpal Singh Gill, Rajkumar Buyya \& Chengzhong Xu, ChainsFormer: A Chain Latency-Aware Resource Provisioning Approach for Microservices Cluster, in: International Conference on Service-Oriented Computing, 2023.
\smallskip\item \label{SP253} Mohit Kumar, Jitendra Kumar Samriya, Kalka Dubey, Sukhpal Singh Gill, QoS-aware resource scheduling using whale optimization algorithm for microservice applications, Wiley Software and Experience 54 (2023).
\smallskip\item \label{SP254} Chunyang Meng, Shijie Song, Haogang Tong, Maolin Pan, Yang Yu, DeepScaler: Holistic Autoscaling for Microservices Based on Spatiotemporal GNN with Adaptive Graph Learnin, in: 38th IEEE/ACM International Conference on Automated Software Engineering (ASE), 2023.
\smallskip\item \label{SP255} Mohamed Samir, Khaled T. Wassif, Soha H. Makady, Proactive Auto-Scaling Approach of Production Applications Using an Ensemble Model, IEEE Access 11 (2023).
\smallskip\item \label{SP256} Duc-Hung LUONG, Huu-Trung THIEU, Abdelkader OUTTAGARTS, Yacine GHAMRI-DOUDANE, Predictive Autoscaling Orchestration for Cloud-native Telecom Microservices, in: IEEE 5G World Forum, 2018.
\smallskip\item \label{SP257} Bing Tang, Xiaoyuan Zhang, Qing Yang, Xin Qi, Fayez Alqahtani, Amr Tolba, Cost-optimized Internet of Things application deployment in edge computing environment, Wiley International Journal of Communication Systems (2023).
\smallskip\item \label{SP258} Yuan Meng, Shenglin Zhang, Yongqian Sun, Ruru Zhang, Zhilong Hu, Yiyin Zhang, Chenyang Jia, Zhaogang Wang, Dan Pei, Localizing Failure Root Causes in a Microservice through Causality Inference, in: 2020 IEEE/ACM 28th International Symposium on Quality of Service (IWQoS), 2020.
\smallskip\item \label{SP259} Carlos Guerrero, Isaac Lera, Carlos Juiz, Resource optimization of container orchestration: a case study in multi-cloud microservices-based applications., The Journal of Supercomputing 74 (2018).
\smallskip\item \label{SP260} Yimeng Wang, Cong Zhao, Shusen Yang, Xuebin Ren, Luhui Wang, Peng Zhao, Xinyu Yang, MPCSM: Microservice Placement for Edge-Cloud Collaborative Smart Manufacturing, IEEE Transactions on Industrial Informatics 17 (2021).
\smallskip\item \label{SP261} Jinjin Lin, Pengfei Chen \& Zibin Zheng, Microscope: Pinpoint performance issues with causal graphs in micro- service environments, in: Service-Oriented Computing. ICSOC 2018, 2018.
\smallskip\item \label{SP262} Adha Hrusto, Emelie Engström, Per Runeson, Towards optimization of anomaly detection in DevOps, Information and Software Technology 160 (2023).
\smallskip\item \label{SP263} Dacheng Zhou, Hongchang Chen, Ke Shang, Guozhen Cheng, Jianpeng Zhang, Hongchao Hu, Cushion: A proactive resource provisioning method to mitigate SLO violations for containerized microservices, IET Communications (2022).
\smallskip\item \label{SP264} Li, N., Tan, Y., Wang, X., Li, B., Luo, J., SCORE: A Resource-Efficient Microservice Orchestration Model Based on Spectral Clustering in Edge Computing, in: Service-Oriented Computing. ICSOC 2022, 2022.
\smallskip\item \label{SP265} Fan Guisheng, Chen Liang, Yu Huiqun, Qi Wei, Multi-objective optimization of container-based microservice scheduling in edge computing, Computer Science and Information Systems 18 (2021).
\smallskip\item \label{SP266} Ma W, Wang R, Gu Y, Meng Q, Huang H, Deng S, Wu Y, Multi-objective microservice deployment optimization via a knowledge-driven evolutionary algorithm, Complex \& Intelligent Systems 7 (2021).
\smallskip\item \label{SP267} Guangba Yu, Pengfei Chen, et al, MicroRank: End-to-end latency issue localization with extended spectrum analysis in microservice environments, in: 2021 World Wide Web Conference, WWW 2021, 2021.
\smallskip\item \label{SP268} F Faticanti, M Savi, F De Pellegrini, Locality-aware deployment of application microservices for multi-domain fog computing, Computer Communications 203 (2023).
\smallskip\item \label{SP269} Ayoub Benayache, Azeddine Bilami, Sami Barkat, Pascal Lorenz, Hafnaoui Taleb, MsM: A microservice middleware for smart WSN-based IoT application, Journal of Network and Computer Applications 111 (2019).
\smallskip\item \label{SP271} CT Joseph, K Chandrasekaran, IntMA: Dynamic interaction-aware resource allocation for containerized microservices in cloud environments, Journal of Systems Architecture (2020).
\smallskip\item \label{SP272} AK Kalia, J Xiao, C Lin, S Sinha, J Rofrano, M Vukovic, D Banerjee, Mono2micro: an ai-based toolchain for evolving monolithic enterprise applications to a microservice architecture, in: ESEC/FSE 2020: Proceedings of the 28th ACM Joint Meeting on European Software Engineering Conference and Symposium on the Foundations of Software Engineering, 2020.
\smallskip\item \label{SP273} De Alwis, A.A.C., Barros, A., Fidge, C., Polyvyanyy, A., Remodularization Analysis for Microservice Discovery Using Syntactic and Semantic Clustering, in: Advanced Information Systems Engineering. CAiSE 2020, 2020.
\end{enumerate}
}



\end{appendices}


\end{document}